\documentclass[12pt,preprint]{aastex}
\newcommand{\U}{\mbox{$u^*$}}
\newcommand{\G}{\mbox{$g'$}}
\newcommand{\R}{\mbox{$r'$}}
\newcommand{\I}{\mbox{$i'$}}
\newcommand{\Z}{\mbox{$z'$}}
\newcommand{\swarp}{\mbox{\tt SWarp}}
\shorttitle{CFHTLS}
\shortauthors{Gwyn}

\begin{document}


\title{The CFHT Legacy Survey: stacked images and catalogs}


\author{Stephen. D. J. Gwyn}
\affil{
Canadian Astronomical Data Centre,
Herzberg Institute of Astrophysics,
5071 West Saanich Road,
Victoria, British Columbia,
Canada   V9E 2E7
}
\email{Stephen.Gwyn@nrc-cnrc.gc.ca}



\begin{abstract}
This paper describes the image stacks and catalogs of the CFHT Legacy
Survey produced using the MegaPipe data pipeline at the Canadian
Astronomy Data Centre.  The Legacy Survey is divided into to two
parts: The Deep Survey consists of 4 fields each of 1 square degree,
with magnitude limits (50\% completeness for point sources) of
$u=27.5$, $g=27.9$, $r=27.7$, $i=27.4$, and $z=26.2$, and contains
$1.6\times 10^6$ sources.  The Wide Survey consists of 150 square
degrees split over 4 fields, with magnitude limits of
$u=26.0$, $g=26.5$, $r=25.9$, $i=25.7$, and $z=24.6$, and contains
$3\times10^7$ sources. 
 This paper describes the calibration, image
stacking and catalog generation process.  The images and catalogs are
available on the web through several interfaces: normal image and text
file catalog downloads, a ``Google Sky'' interface, an image cutout
service, and a catalog database query service.
\end{abstract}




\section{INTRODUCTION}
\label{sec:intro}

The Canada France Hawaii Telescope Legacy Survey (CFHTLS) consisted of
three surveys: The Deep Survey imaged 4 square degrees; the primary
goal was to measure cosmological parameters using Type Ia supernovae
as standard candles. This portion of the CFHTLS is also known as the
Super Nova Legacy Survey \cite[SNLS]{SNLS1}. The Wide survey covered a
wider area to shallower depth. It consists of 171 overlapping
pointings.  which (after accounting for overlaps) cover 150 square
degrees.  Weak lensing and large scale structure studies were the
primary science goals.  The Very Wide portion of the survey covered
the ecliptic plane and was designed to study Kuiper Belt Objects.  It
was cancelled mid-way through the project but has become the
Canada-France Ecliptic Plane Survey \cite[CFEPS]{cfeps2009}.

Interestingly, image stacks are secondary to the main science goals of
the CFHTLS.  The main reason for the Deep Fields is supernova
cosmology.  The supernovae are discovered using stacks generated on a
nightly basis. Deeper stacks are generated but only as a reference for
difference imaging. Similarly, the main science driver of the Wide
fields is weak lensing. The actual weak lensing measurements are done
on individual images in order to accurately quantify the PSF. The 
value of the Very Wide survey, which was looking for moving objects,
is also in the individual images.

Now that the primary science goals of the CFHTLS are well in hand,
the archival data becomes important. Creating stacks and catalogs
greatly increases the usefulness of archival MegaCam data.

MegaPipe \citep{gwyn2008} has been used to process over 2000 square
degrees of MegaCam imaging over the last four years.  Twice a year it
goes through the CFHT archive at the CADC, determines which images are
worth combining, computes accurate astrometric/photometric
calibrations, stacks the images, and makes the stacked images to the
astronomical community. In addition, it is used process PI data (on
request) and to process data from large surveys such as the Next
Generation Virgo Survey (Ferrarese et al. in preparation)

This paper describes image stacks and catalogs of the CFHTLS Deep and
Wide surveys generated by the MegaPipe image processing pipeline.  It
presents the astrometric and photometric calibration, the image
stacking and catalog generation, and quality control.  This paper
outlines the various web applications that astronomers can use to
download the data. This paper however does not describe the stacking
of the Very Wide data; everything worth stacking was stacked as part
of regular semi-annual MegaPipe processing.

Some of the methods presented in this paper have already been
described in the main MegaPipe paper \citep{gwyn2008}.  For
completeness, this paper describes these methods again (usually in
less detail), but more emphasis is placed on methods that are specific
to the CFHTLS stack generation and quality control.

The AB magnitude \citep{abmag} system is used throughout this paper.

\section{FIELD LOCATIONS}
\label{sec:field}

\begin{figure}
\plotone{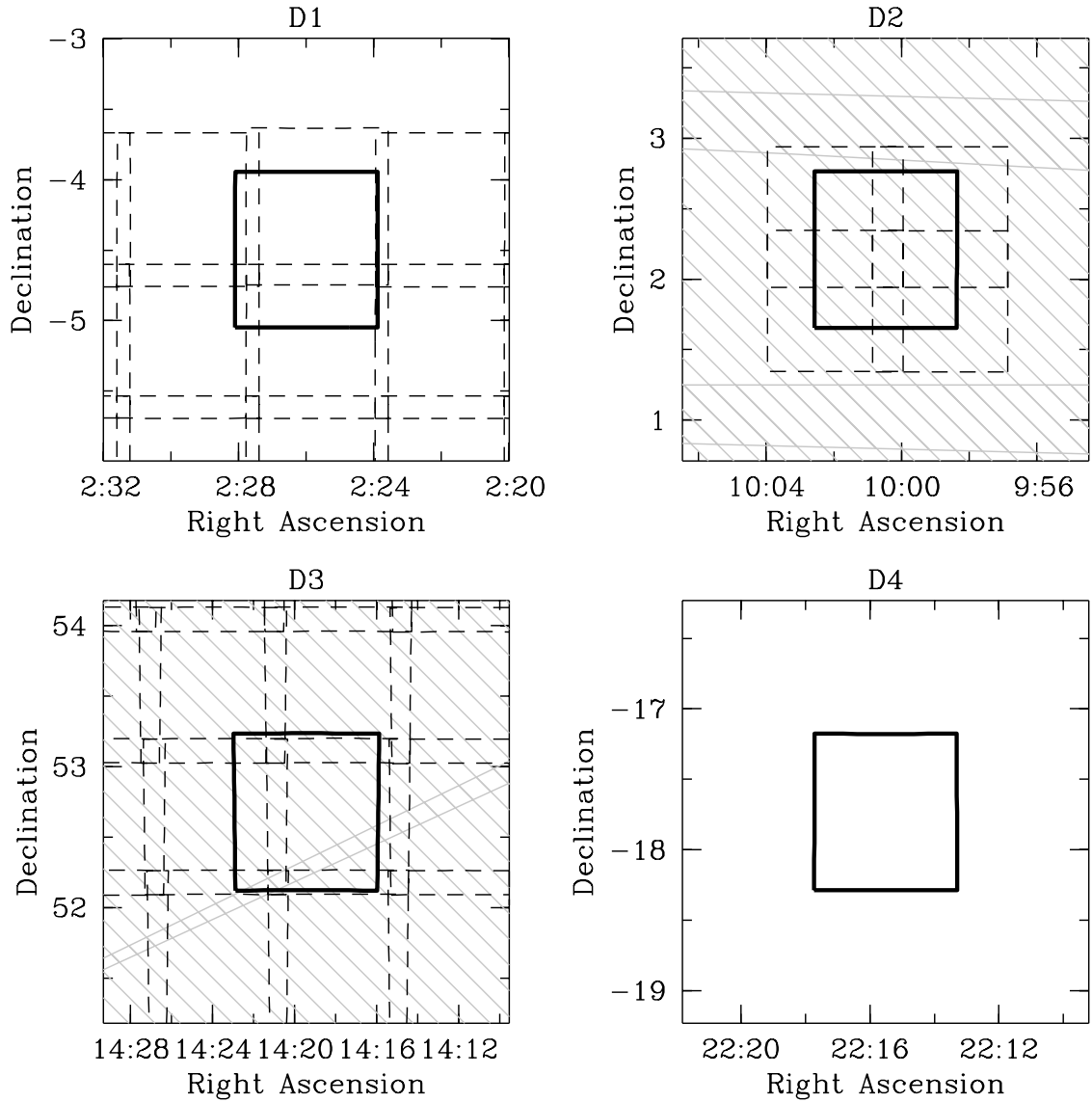}
\caption{
Layout of the CFHTLS Deep fields. The solid boxes show the four
MegaCam pointings. The grey hashed area shows the location of the SDSS.
The dashed boxes in the D1 and D3 panels show the location of the MegaCam W1 and W3 pointings respectively.
The dashed boxes in the D2 panel show the location of the MegaCam COSMOS pointings.
}
\label{fig:deeplaybw}
\end{figure}

The layout and positions of the Deep fields are shown in Figure
\ref{fig:deeplaybw}.  The figure shows the Deep fields themselves as solid black
outlines, with dashed outlines denoting overlapping MegaCam
imaging. The D1 and D3 fields overlap the corresponding Wide fields
and the D2 overlaps the COSMOS \citep{cosmos} \U-band imaging, as
shown.  The grey lines and hashed areas show the positions of the
Sloan Digital Sky Survey \citep[SDSS]{dr7} swaths; the D2 and D3 lie
within the SDSS.

The D1 field lies within the XMM-LSS \citep{xmmlss} and VVDS
\citep{vvdsdeep} fields.  Surveys at other wavelengths also overlap:
near-infrared \citep{VIRMOSdeepIR}, infrared \citep{swireD1}, and
radio \citep{VIRMOSradio}.  In addition, there is a considerable
amount of spectroscopy \citep{VIRMOSspec}

The D2 field lies within the COSMOS survey. There are X-ray
\citep{xCOSMOS}, UV \citep{uvCOSMOS}, infrared \citep{sCOSMOS}, near
IR \citep{imageCOSMOS}, and radio \citep{radioCOSMOS} data available.
In addition, the zCOSMOS survey \citep{zCOSMOS} provides
spectroscopic coverage.

The D3 field overlaps the Groth strip. It is the site of several
previous surveys, including the CFRS \citep{cfrs1} and a LBG survey
\citep{steidel2003}.  The ongoing AEGIS survey also overlaps the D3
\citep{aegis}. It includes HST imaging, X-ray \citep{aegisX}, UV data
from GALEX near-IR \citep{aegisIR} and infrared data from Spitzer, as
well as several thousand galaxy spectra.

\begin{figure}
\plotone{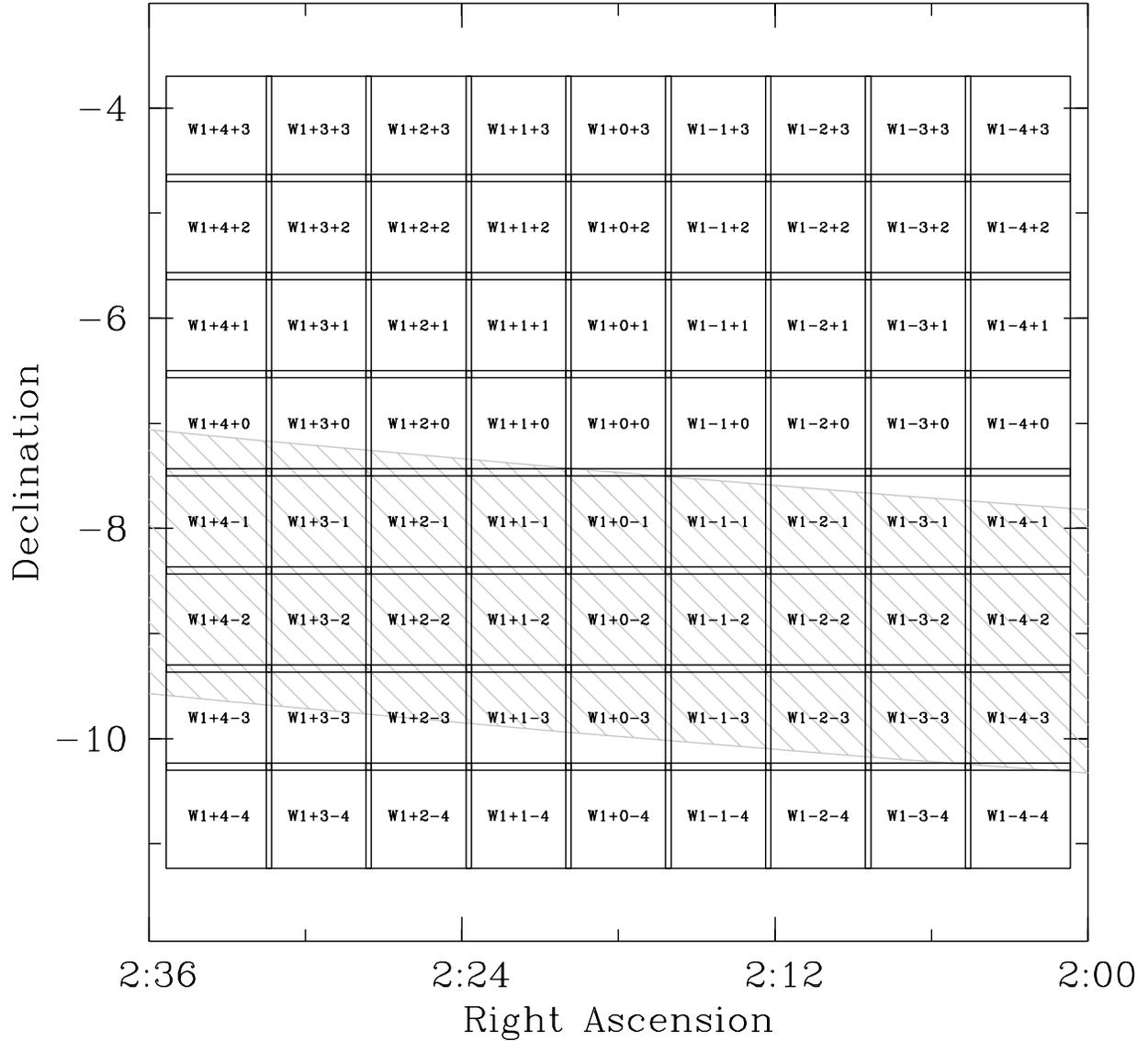}
\caption{
Layout of the CFHTLS W1 field. The labelled boxes show the
MegaCam pointings. The grey hashed area shows the location of the SDSS swath,
which lies across the middle of the field, overlapping several pointings.
}
\label{fig:W1.bw.lay}
\end{figure}

\begin{figure}
\plotone{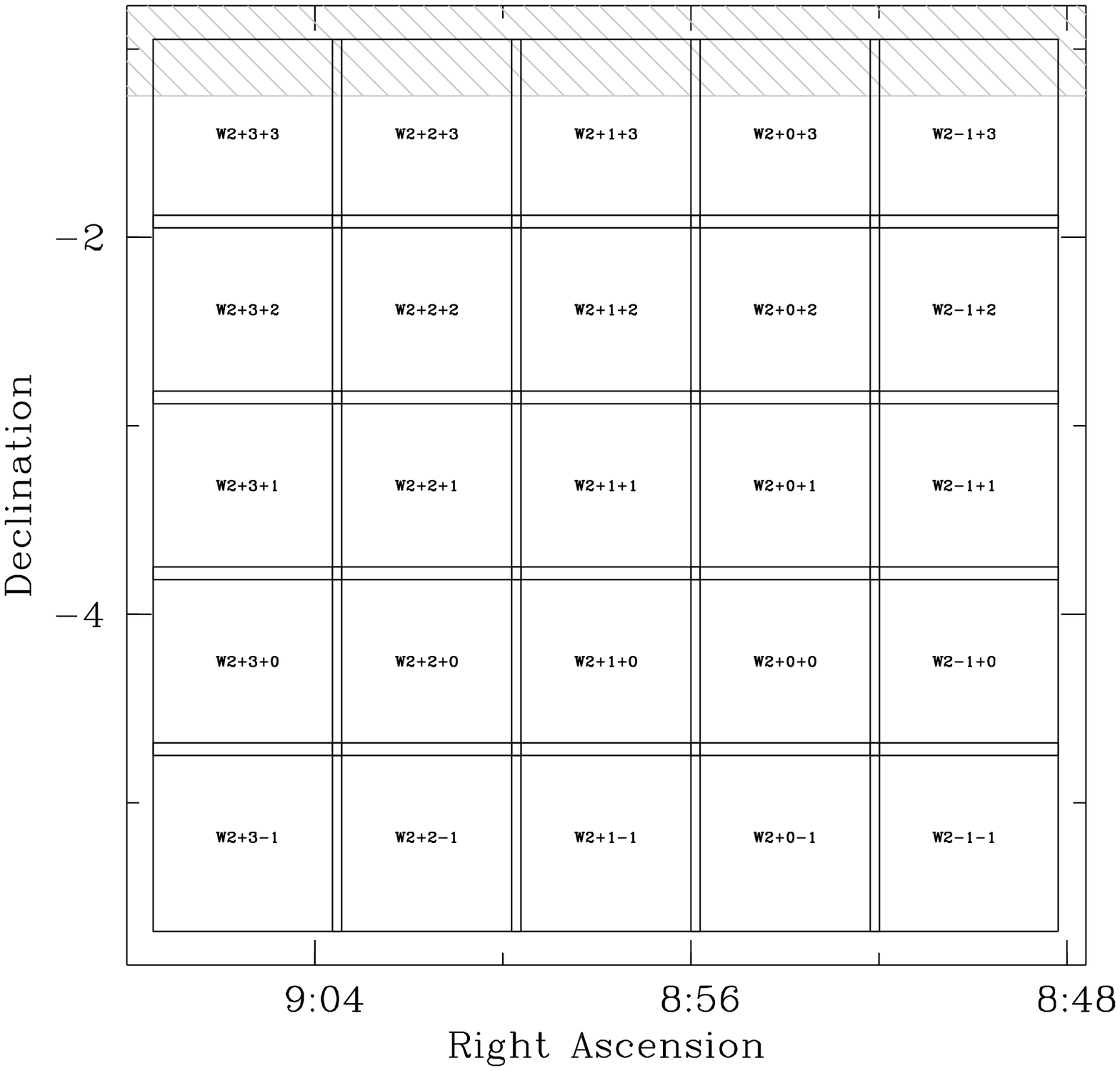}
\caption{
Layout of the CFHTLS W2 field. The labelled  boxes show the
MegaCam pointings. The grey hashed area shows the location of the SDSS swath,
which barely overlaps the field.
}
\label{fig:W2.bw.lay}
\end{figure}

\begin{figure}
\plotone{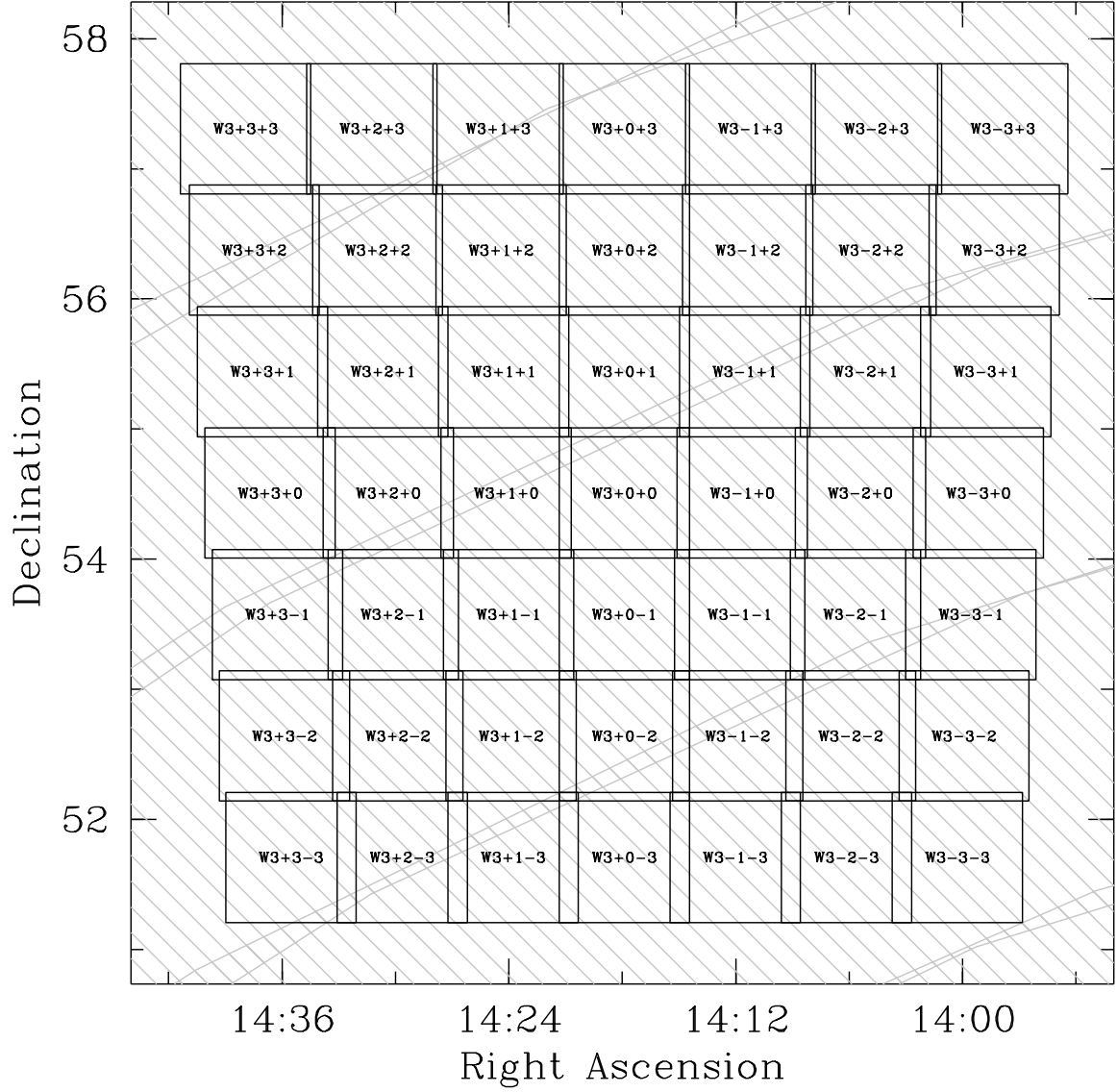}
\caption{
Layout of the CFHTLS W3 field. The labelled boxes show the
MegaCam pointings. The W3 field lies completely in the SDSS as shown by
the grey hashed area.
}
\label{fig:W3.bw.lay}
\end{figure}

\begin{figure}
\plotone{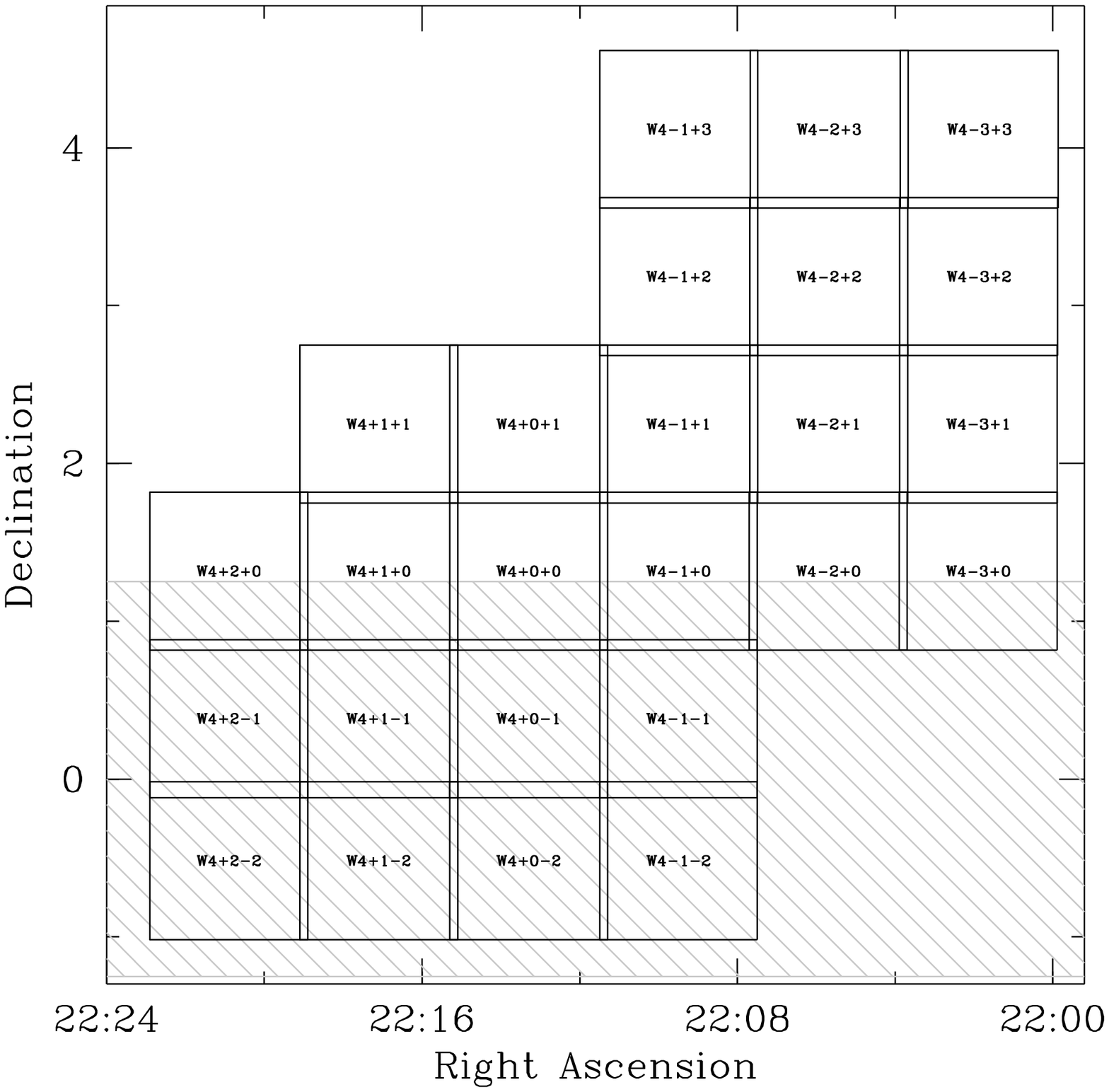}
\caption{
Layout of the CFHTLS W4 field. The labelled boxes show the
MegaCam pointings.  The grey hashed area shows the location of the SDSS swath
which overlaps the field.
}
\label{fig:W4.bw.lay}
\end{figure}

The layout and positions of the Wide fields are shown in Figures
\ref{fig:W1.bw.lay}, \ref{fig:W2.bw.lay}, \ref{fig:W3.bw.lay}, and
\ref{fig:W4.bw.lay}.  The pointings are shown as solid black lines and
the grey lines and hashed areas show the positions of the SDSS
swaths. Some pointings of all four Wide fields overlap the SDSS.  The
W1 and W3 overlap the D1 and D3 respectively.  The W4 lies on the
special area SA22 which is the location of the VVDS 22h field as well
as one of the UKIDSS DXS fields \citep{ukidssGen}.

\section{INPUT DATA}
\label{sec:input}

\subsection{The MegaCam Camera and filters}

The MegaCam camera is a wide field camera mounted at the prime focus
of the Canada France Hawai'i Telescope. It is a mosaic of 36 CCDs,
each $2048\times4612$ pixels, arranged in a $9\times4$ grid.  The
field of view is just under 1 degree on a side. The resolution of
MegaCam is 0.187 arcseconds per pixel. It is located behind
the a wide field corrector and an image stabilizing unit
For further details about the MegaCam/MegaPrime camera, the reader
is referred to \citet{megacam}.

\begin{figure}
\plotone{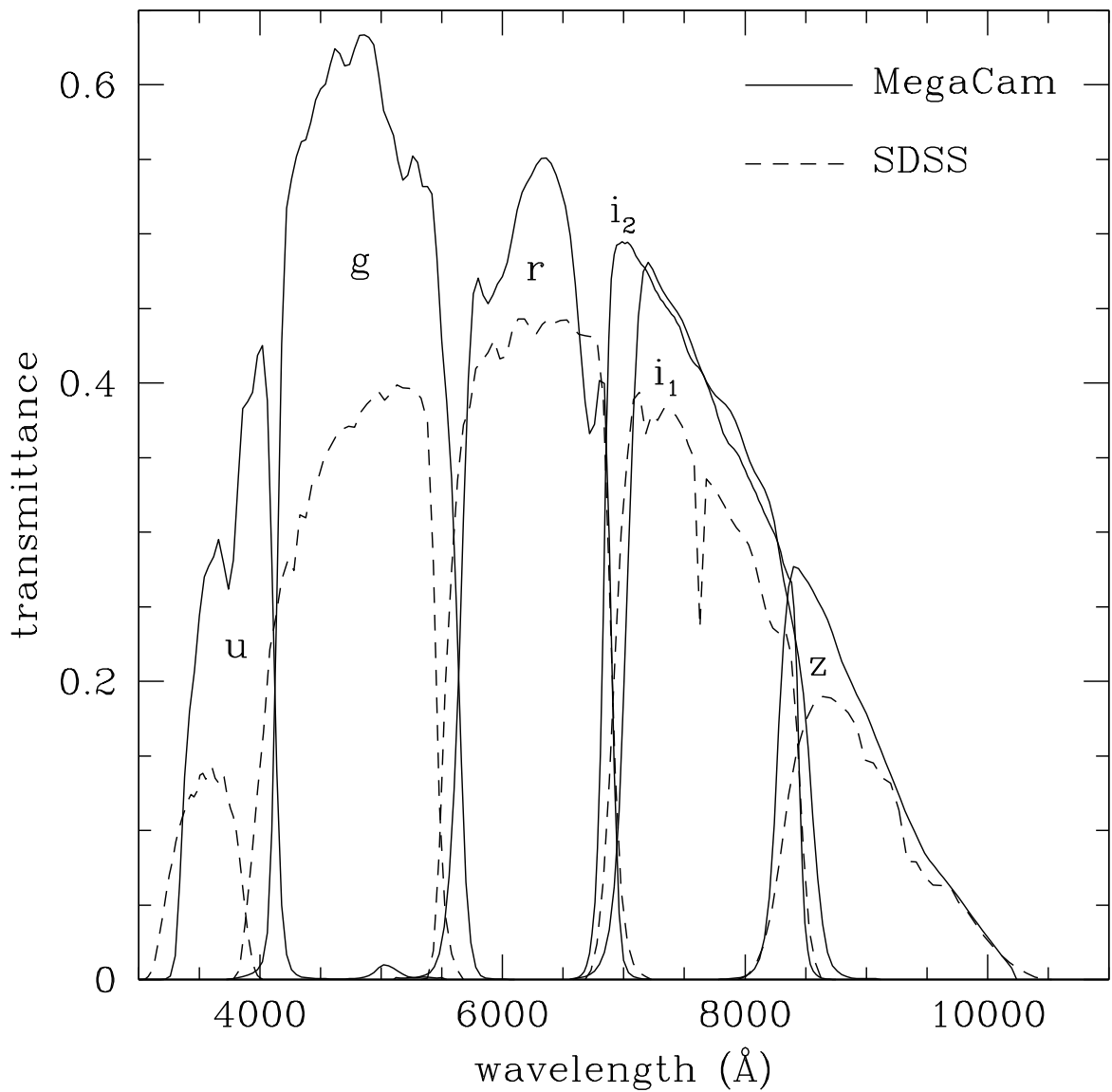}
\caption{Comparing the MegaCam (solid lines) and SDSS (dashed lines)
\U\G\R\I\Z\ filter sets. The transmittance curves show the final
throughput including the filters, the optics and the CCD
response. There are two \I\ filters.  The older one (CFHT filter
i.MP9701) is labelled i1; the newer one  (CFHT filter i.MP9702) is
labelled i2. Note that the new \I\ filter is slightly bluer than the old one.
}
\label{fig:megasdssbw}
\end{figure}

MegaCam is equipped with 5 broadband filters: \U\G\R\I\Z. The
bandpasses are shown in Figure \ref{fig:megasdssbw}. Halfway through
the survey the \I\ filter was damaged. It was replaced with a new
filter, with a slightly bluer bandpass than the original. The old and
new filters are labeled $i1$ and $i2$ respectively in the figure.

\subsection{Elixir Preprocessing}

All the CFHTLS images were preprocessed using the Elixir pipeline at
CFHT before delivery to the CADC. For full details on the Elixir
pipeline, the reader is referred to \citet{elixir} and the CFHT Elixir
webpage\footnote{http://www.cfht.hawaii.edu/Science/CFHTLS-DATA/dataprocessing.html}
What follows is a summary.

Elixir first applies an overscan correction for each amplifier of each
CCD (two amplifiers per CCD). The bias frame is built of from 20-30
frames. The bias frame is built once per observing run. No dark
current corrections are applied.

Flatfielding is done using twilight flats. These twilight flats are
modified using a photometric superflat to correct for scattered light
and changes in the pixel scale and optical transmission across the
field of the field of view of the detector. The flats are generated
on a run-by-run basis.

For the \I and \Z images only, fringes are removed. A master fringe is
constructed from a large number of images, suitably scaled. The master
fringe image is scaled and subtracted from each image.

An astrometric calibration is done by using the known plate scale and
image orientation to get an approximate plate solution. This is then
refined using the USNO catalog \citep{usno}. The final solution is
typically good to better than an arcsecond.

Elixir does a photometric calibration based on observations of Smith
et al. \citep{smith} standard stars taken at the beginning and end of
each night.  The accuracy and necessary corrections to this
calibration are discussed in section \ref{ssec:photomElixir}.

Elixir provides a pixel mask for MegaCam images indicating
the location of bad pixels and columns. It is used in the
stacking procedure described in Section \ref{sec:stack}.

\subsection{Input Image Selection}

The images were selected from the CFHT archive at the Canadian
Astronomy Data Centre. All images with centers within 0.05 degrees of
the nominal pointing centers were included. For the most part these
are CFHTLS images, however this also includes a number of public
images from other programs. The COSMOS field consists of 4 MegaCam
pointings arranged around the D2 field, and centered on it.  The
\U-band images from the COSMOS program were included. Each of these
images covers about a quarter of the D2 field.  Images taken in the
new \I\ filter were not included in the Deep field stacks.  The complete
list of CFHT exposures which were included in the stacks is available
on-line \footnote{\tt
  http://www.cadc.hia.nrc.gc.ca/community/CFHTLS-SG/docs/cfhtlsinput.html}.

\subsection{Input Image Quality Control}

The images were checked in a number of ways before being included in
the stacks.  SExtractor was run on each image. The stellar locus was
identified automatically. The properties of objects in the stellar
locus were used to set rejection criteria.

\begin{figure}
\plotone{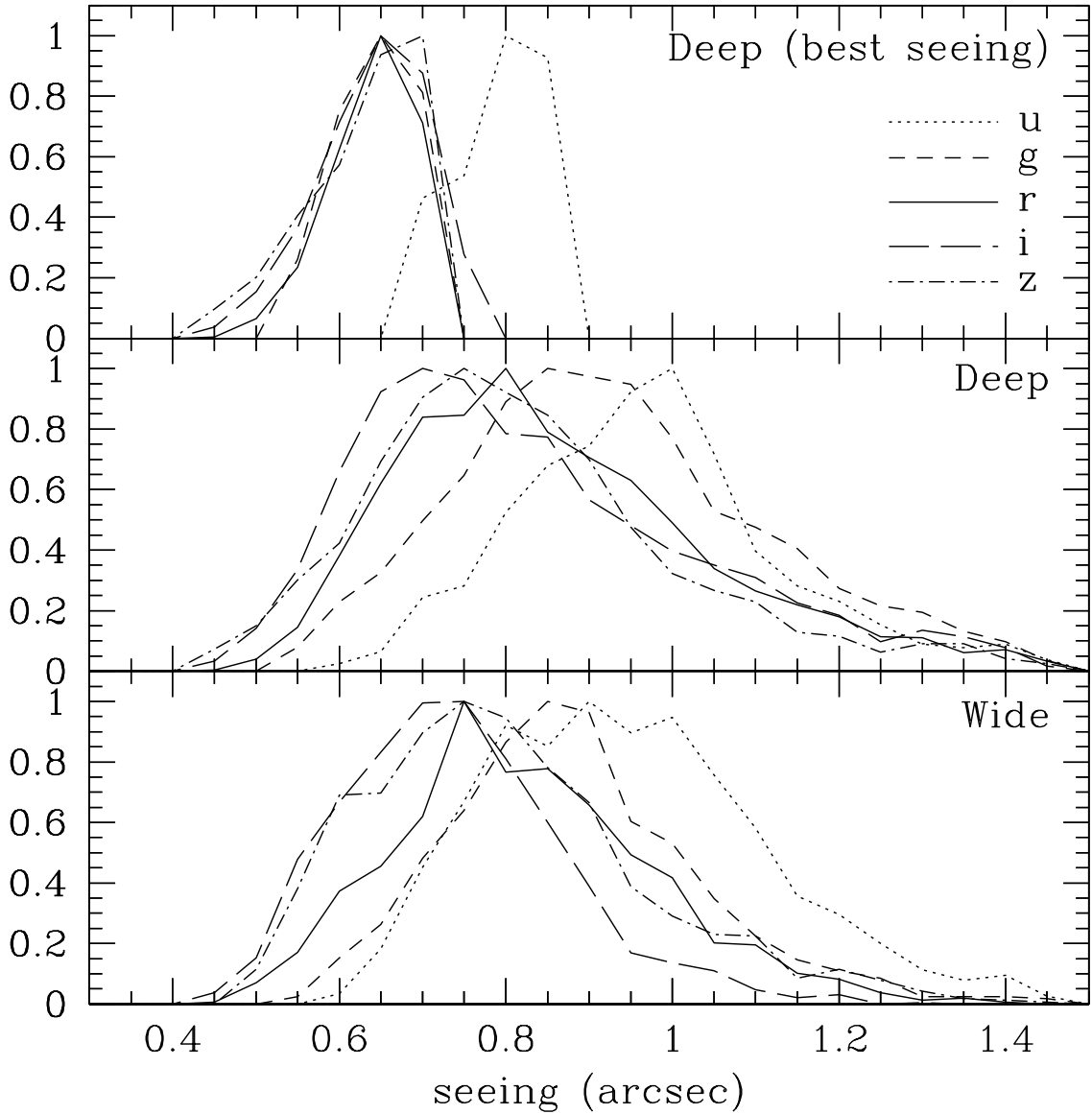}
\caption{ 
Image quality of the CFHTLS input images.  The seeing distribution,
split by filter, is shown for the different surveys.  As discussed in
the text, two versions of the Deep survey were generated.  The
uppermost panel shows the seeing distribution of the input image
``good seeing'' stacks.  The median of the \G\R\I\Z\ image quality is
0.65$''$.  The median of the \U\ image quality is 0.8$''$. The
distributions are truncated. The middle panel shows the seeing
distribution of the input images to the ``full'' stacks. The
bottom panel shows the seeing distribution for the Wide survey.
}
\label{fig:cfhtlsiqdistbw}
\end{figure}

The image quality (ie, seeing) was measured. If the seeing was worse
than 1.5$''$,  the image was rejected. Figure \ref{fig:cfhtlsiqdistbw}
shows the distribution of seeing for the CFHTLS input images
split by survey and filter.

In addition, if the PSF is noticeably trailed (due to poor tracking),
the image was rejected.  If $a$ and $b$ are the semi-major and
semi-minor axes of sources in the stellar locus,
the asymmetry index, $A$, is given by
\begin{equation}
A=1-b/a
\end{equation}
The average of the this index over each field, $\overline A$, is always slightly
greater than 0 due to the optical aberrations in the MegaPrime
optics. This index is crude, but is more than sufficient to rejected
trailed images.  Images with an average asymmetry index $\overline A >0.2$ are
noticeably trailed and were rejected.

Images taken in conditions of poor transparency were also rejected. If
the measured zero-point was more than 1 magnitude below the nominal
value, the image was rejected.

Finally, a subsection of one chip of each image was examined by eye
for other abnormalities.  For example, one image flaw which cannot be
detected automatically is a discrete jump in telescope position during
the exposure. In this case, each source is doubled (and bright stars
will leave a trail between the two image) but the PSF will stay the
same. This is immediately obvious to eye, but hard to detect
automatically.

\subsection{Best Seeing Image Selection}

Two versions of the Deep field stacks were produced: the ``full''
stacks and the ``best seeing'' stacks. This section describes how the
images that went into the ``best seeing'' images were selected.

When making ``best seeing'' stacks, one must decide how many input
images to include and how many to throw away. There is trade-off
between limiting magnitude and image quality (IQ). As one includes
more images with increasingly bad image quality, the output image
quality degrades while the depth increases.

Quantitatively, the IQ of a stack of $N$ images is very well estimated by the median of
the IQ of its input images, i.e., the IQ of input image $N/2$. To estimate
the limiting magnitude, two assumptions were made: 1) the effects of
IQ on depth are not important (not quite true); and 2) all the input
images have the same exposure time (for the CFHTLS Deep Fields, quite
true). In this simplified case, the limiting magnitude of a stack
of N images goes approximately as $1.25 \log_{10}N$.

\begin{figure}
\plotone{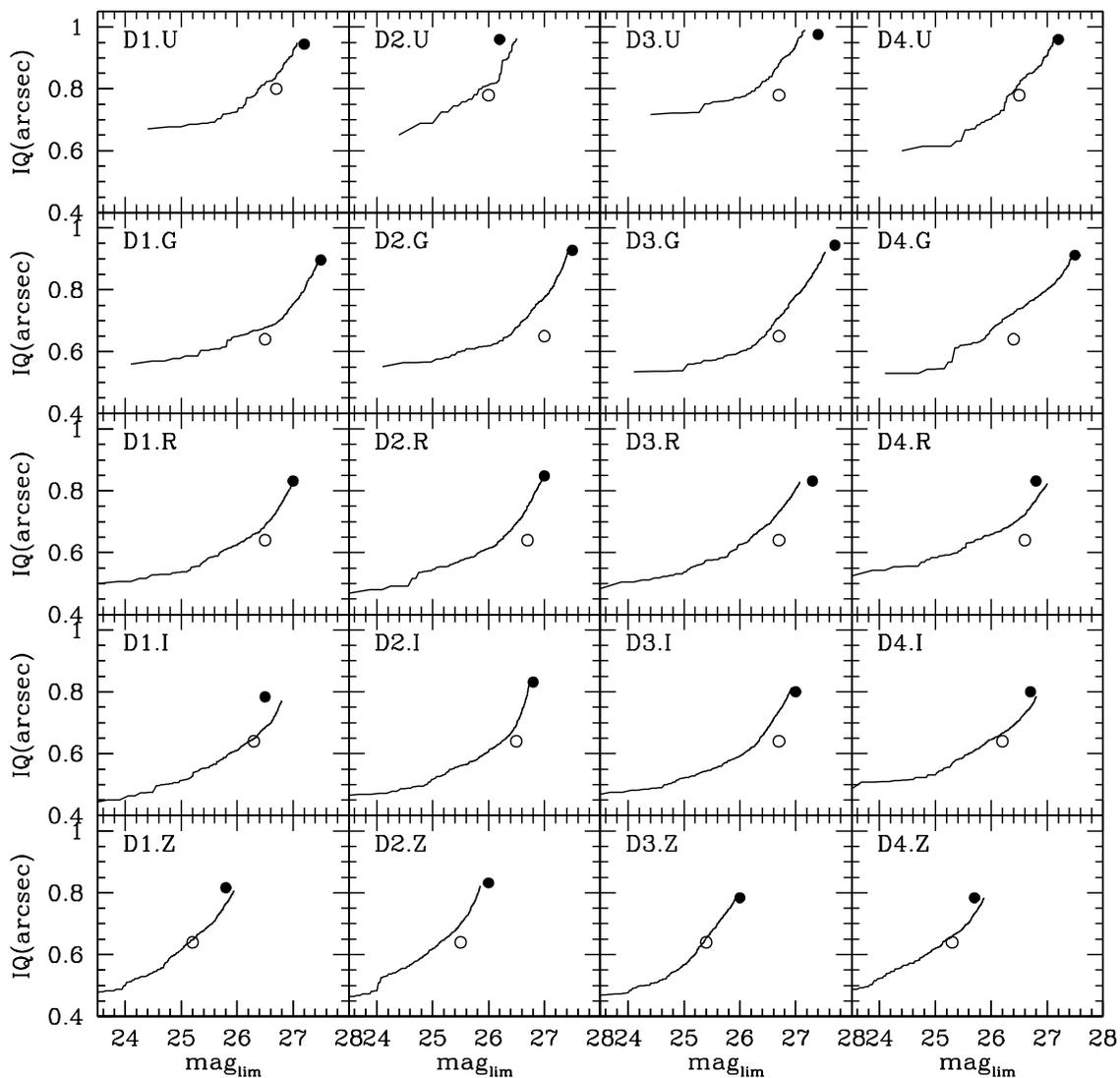}
\caption{ 
Seeing vs limiting magnitude for the CFHTLS Deep fields.  If the input
images are sorted by image quality, as more images are stacked, the
output image quality should become worse but the limiting
magnitude should become deeper, as shown by the black line.  The solid
points show the actual locus of the ``full'' stacks.  The open dots
show the actual locus of the ``best seeing'' stacks.
}
\label{fig:fiqbw}
\end{figure}

These two measures were applied to the input images of the CFHTLS Deep
fields. The images were sorted in order of image quality: best seeing
first, worst seeing last. As the number of input images increases, the
image quality of the stack will get worse, but the limiting magnitude
will get deeper.  The results are shown in the grid of plots of Figure
\ref{fig:fiqbw}. The expected image quality is plotted against expected
limiting magnitude for each the 5 filters and 4 deep fields. The black
lines show both IQ and limiting magnitude increasing with number of
images.  The red points show the locus of the "full" stacks.

It can be seen that while the median of the image quality of the input
images is an excellent predictor of the IQ of the output stacks, the
limiting magnitude prediction is less accurate. Deviations of several
tenth of a magnitude from the prediction are not uncommon. This not
surprising, given the simplicity of the underlying assumptions.

The question becomes: is there a point on the graphs where an
improvement in image quality comes at only a small cost in limiting
magnitude? Some of the graphs exhibit kinks where the slope of the
line changes. Obviously, it is better to be at or slightly below these
kinks. Ideally all the output images would have the same IQ, to
increase the accuracy of matched aperture photometry.

Instead of picking a single threshold (for example stacking the best
25\% of the images, or stacking only images with IQ=0.7$''$ or better),
images were included to produce a desired output image quality. This
is possible because, as shown by the open dots on the Figure \ref{fig:fiqbw}, the
median input IQ is an excellent predictor of the output IQ. Therefore,
when choosing input images, one first sorts the images by increasing
IQ. Then one goes down the ordered list until an input image with the
target IQ is found, and then selects twice that number of images. For
example, if one wants a 0.65$''$ seeing stack, and there are 41 images
with seeing better than 0.65$''$, one should stack the best 82 images.

The chosen target image quality was 0.65$''$ for the \G\R\I\Z\ filters and
0.8$''$ for the \U\ filter. Decreasing the seeing below 0.8 becomes
rapidly prohibitive for the \U-band. Similarly, the \Z- and particular
the \G-band depths decrease rapidly if the target IQ is decreased even
slightly. The \U-band target is different than the other bands because
targeting IQ=0.65$''$ would include no images, and targeting IQ=0.8$''$ in
the other bands would improve the seeing only slightly relative to the
"full" stacks. 

These criteria were applied to the input images of the CFHTLS Deep
Fields and the chosen images were stacked. The resulting image quality
and limiting magnitudes are plotted as filled dots on Figure
\ref{fig:fiqbw}. The image quality is consistently on (or very slightly
below) the prediction. The limiting magnitudes are usually better than
predicted, since the prediction relies on the (not completely correct)
assumption that IQ does not affect depth.

\section{ASTROMETRIC CALIBRATION}
\label{sec:astrom}

The astrometric calibration consisted of generating a catalog of
sources from the input MegaCam images, retrieving an astrometric
reference catalog, and determining the mapping between the two.  The
mapping was then expressed in terms the FITS WCS (World Coordinate System)
parameters.  This mapping was then used by SWarp in the resampling phase of
the stacking procedure.

\subsection{Source Catalog Generation}
\label{ssec:astrosource}
SExtractor \citep{hihi} was used to generate the image source catalog.
The source catalog was limited to brighter sources with well defined
centres. The parameters were set such that sources must have 5
continuous pixels 5 $\sigma$ above the sky level in order to count as
a detection.  ({\tt DETECT\underline{~}MINAREA}=5 and {\tt
DETECT\underline{~}THRESH}=5 in SExtractor parlance).  The catalog
was cleaned. Cosmic rays were rejected by removing sources which are
more compact than stars as measured by their half-light radius.
Extended sources were identified and rejected as those sources whose
Kron magnitudes were more than 2 magnitudes greater than their
1.5\arcsec\ radius magnitudes.

\subsection{Reference Catalogs}
\label{ssec:astroref}

A bootstrapping series of astrometric reference catalogs were used.
The Naval Observatory Merged Astrometric Dataset \cite[NOMAD]{nomad}
was used a the first external astrometric catalog.  NOMAD has a higher
source density and slightly better astrometric accuracy than the USNO
catalog used by Elixir.  Where available, the SDSS DR7 \citep{dr7} was
used to refine the astrometry.  The final reference catalog was
generated from a subset of the CFHTLS images themselves.  For the Deep
fields the a subset of the images whose centers were the furthest away
from field center were used. For the D2 field, some archival COSMOS
images were included; for the D3, some W3 images were included. Only
\I-band images were used.  For the Wide fields, images from the main
pointings were supplemented with images from the ``pre-imaging''
survey. Here, \R\ band images were used.  Each of these images were
calibrated using the NOMAD/SDSS astrometric solutions. Source catalogs
were generated for each images with SExtractor; the catalogs were
cleaned as noted in Section \ref{ssec:astrosource} with a further
magnitude cut at mag=23..  The ($x,y$) positions were converted to
(RA, Dec) using the astrometric solutions.  The catalogs for a given
survey and field were then merged, forming an final reference catalog
for each field. These internal catalogs have smaller astrometric
residuals and a higher source density then the external
catalogs. Using these internal catalogs left the external astrometric
accuracy unchanged (or improved very slightly) but dramatically
improved the internal astrometric accuracy.

\subsection{Mapping}
\label{ssec:astromap}

The first step of the mapping was the initial matching up of the
observed catalogs (section \ref{ssec:astrosource}) to the
reference catalogs (section \ref{ssec:astroref}).  For the
matching, each CCD of the mosaic was treated independently.  The ($x,y$)
coordinates of the observed catalog were converted to (RA, Dec) using
the initial Elixir WCS. The catalogs were shifted in RA and Dec with
respect to one another until the best match between the two catalogs
was found. This simple up-down/left-right shift method was found to be
sufficient. More complex methods such as triangle matching or Fourier
transform methods were not necessary since the plate scale and image
orientation were known beforehand.  If there was no good match for a
particular CCD (for example, when the initial Elixir WCS unusually
erroneous), its WCS was replaced with a default WCS and the matching
procedure was restarted. Once the matching was complete, the
astrometric fitting could begin. The initial matching was always done
with the NOMAD catalog; the higher source density of the SDSS and the
internal catalogs would have made  matching prohibitively slow.
Typically 20 to 50 sources per CCD were found with this initial
matching.

Once the initial match was complete, the first astrometric solution
was determined. Using this astrometric
solution, the matching was restarted, this time with tighter
constraints on what constituted a good match between a observed source
and an source in the reference catalogs. This procedure was repeated
several times, steadily shrinking the matching radius from 8
arcseconds to 0.5 arcseconds. The order of astrometric solution
increased from a linear fit in the first iteration to a $3^{\rm rd}$
order solution by the last. This match/map/repeat procedure was
done for each CCD individually.

The final astrometric mapping for each image was determined on the
scale of the whole mosaic, all 36 CCDs at once. This final mapping
consists of a linear mapping for each CCD: shift and skew for each
axis. In terms of FITS WCS keywords, this corresponds to the
CRVAL/CRPIX keywords and the CD matrix. The radial distortion
correction is modeled as:
\begin{equation}
R=r (1 + a_2 r^2 + a_4 r^4),
\end{equation}
where $R$ is the true radius and $r$ is the measured radius.  This
equation is analogous to equation 3 of \cite{chiu76} and the equation
in Section 3.1 of \cite{mocam}.  Thus, only 2 parameters were used to
describe the distortion for the whole mosaic. In other systems, a
second or third order function is needed to describe the distortion,
requiring up to 20 parameters per CCD, or a total of 720 parameters
per mosaic. The best values of $a_2$ and $a_4$ were found through a
crude but robust mapping of parameter space.  Once determined, the
astrometric mapping described by the values $a_2$ and $a_4$ was
re-expressed in terms of the FITS WCS {\tt PVn\underline{~}n} keywords
\citep{wcs3}, so as to be understood by SWarp.

The internal and external astrometric uncertainties are estimated to be
0.04$''$ and 0.07$''$ respectively, as discussed in section \ref{ssec:qastrom}.

\section{PHOTOMETRIC CALIBRATION}
\label{sec:photom}

The SDSS DR7 \citep{dr7} served as the basis of the photometric
calibration. For images lying within the SDSS, the SDSS is used as
in-field standards, as described in section \ref{ssec:photomSDSS}.
Outside the SDSS, images taken on photometric nights were calibrated
using the Elixir calibration, but modified by comparing the Elixir
zero-points to other images taken on the same night that did lie
within the SDSS, as described in \ref{ssec:photomElixir}.  Images
taken on non-photometric nights, lying outside the SDSS were
calibrated by bootstrapping from overlapping images calibrated by one
of the other two methods, as described in section
\ref{ssec:photomNeigh}.

\subsection{Using SDSS}
\label{ssec:photomSDSS}

The Sloan $ugriz$ filters are not identical to the
MegaCam filters, as shown in Figure \ref{fig:megasdssbw}. The color
terms between the two filter sets can be described by the following
equations:
\begin{eqnarray}
\label{eqn:trans}
  \U &=& u_{SDSS}  -0.211  \times (u_{SDSS} - g_{SDSS})\nonumber \\
  \G &=& g_{SDSS}  -0.155  \times (g_{SDSS} - r_{SDSS})\nonumber \\
  \R &=& r_{SDSS}  -0.030  \times (r_{SDSS} - i_{SDSS})\nonumber \\
  \I &=& i_{SDSS}  -0.102  \times (r_{SDSS} - i_{SDSS})\nonumber \\
  \Z &=& z_{SDSS}  +0.036  \times (i_{SDSS} - z_{SDSS})
\end{eqnarray}
after \cite{regnault}.

All images lying completely in the SDSS were directly calibrated without
referring to other standard stars such as \cite{smith}
standards. The systematics in the SDSS photometry are about 0.02
magnitudes \citep{ivezic2004} The presence of at least 1000 usable
sources in each square degree reduced the random error to effectively
zero. It was possible to calibrate the CCDs of the mosaic
individually with about 30 standards in each.  For each MegaCam image,
the corresponding catalog (in instrumental magnitudes) was matched to
the to SDSS catalog for that patch of sky.  The difference between the
instrumental MegaCam magnitudes and the SDSS magnitudes gave the
zero-point for that exposure or that CCD.  The zero-point was
determined by median.

There are about $10^4$ SDSS sources per square degree, but when one cuts by
stellarity and magnitude this number drops to around 1000.  Only stars
were used since the above color terms are more appropriate to stars
than galaxies.  Only objects with $17<$mag$<20$ were used; the brighter
objects are usually saturated in the MegaCam image and including the
fainter objects tends to only increase the noise in the median.  This
process could be used any night; it was not necessary for the night to
be photometric. The D2, D3, and the W3 were calibrated in this manner.

\subsection{Using Elixir}
\label{ssec:photomElixir}

For pointings outside the SDSS, the Elixir photometric keywords were
used, with modifications.  The Elixir zero-points were compared to
those determined from the SDSS using the procedure above for a large
number of archival MegaCam images.  There are systematic offsets
between the two sets of zero-points, particularly for the
\U-band. These offsets show variations with epoch, which are caused by
modifications to Elixir pipeline (Cuillandre, private
communication). There are also differential offsets between the CCDs of a
single image. For MegaPipe, the offsets are applied from the Elixir
zero-points to bring them in line with the SDSS zero-points.

The differential offsets are the offsets between different CCDs of a
single MegaCam image, regardless of whether the image was taken
under photometric conditions or not. Ideally, the photometric zero-point should
be the same across the MegaCam field. In practice, differences exist.
To investigate the differential zero-point offsets, the MegaCam
archive was searched for images matching the following criteria:

\begin{itemize}
\item Each image must have an exposure time greater than 60 seconds
\item Each image must lie within the SDSS
\item No two images from the same camera run should lie with 0.3 degrees
  of each other. 
\end{itemize}
The last condition is to ensure that the isolated patches of bad SDSS
photometry don't affect the final result..

Each of the 12 000 images matching these criteria was astrometrically
calibrated using the standard MegaPipe pipeline. The photometric
zero-points were computed separately for each CCD. The difference
between the individual zero-points and the average of all the
zero-points was computed. Since the images were not necessarily taken
under photometric conditions, the Elixir zero-point may or may not be
valid, and therefore was ignored at this stage. The old \I\ filter
(I.MP9701) and the new \I\ filter (I.MP9702) were treated separately.
Note that many of these images (indeed, the majority) are not CFHTLS
images, but archival MegaCam images taken for variety of projects.

\begin{figure}
\plotone{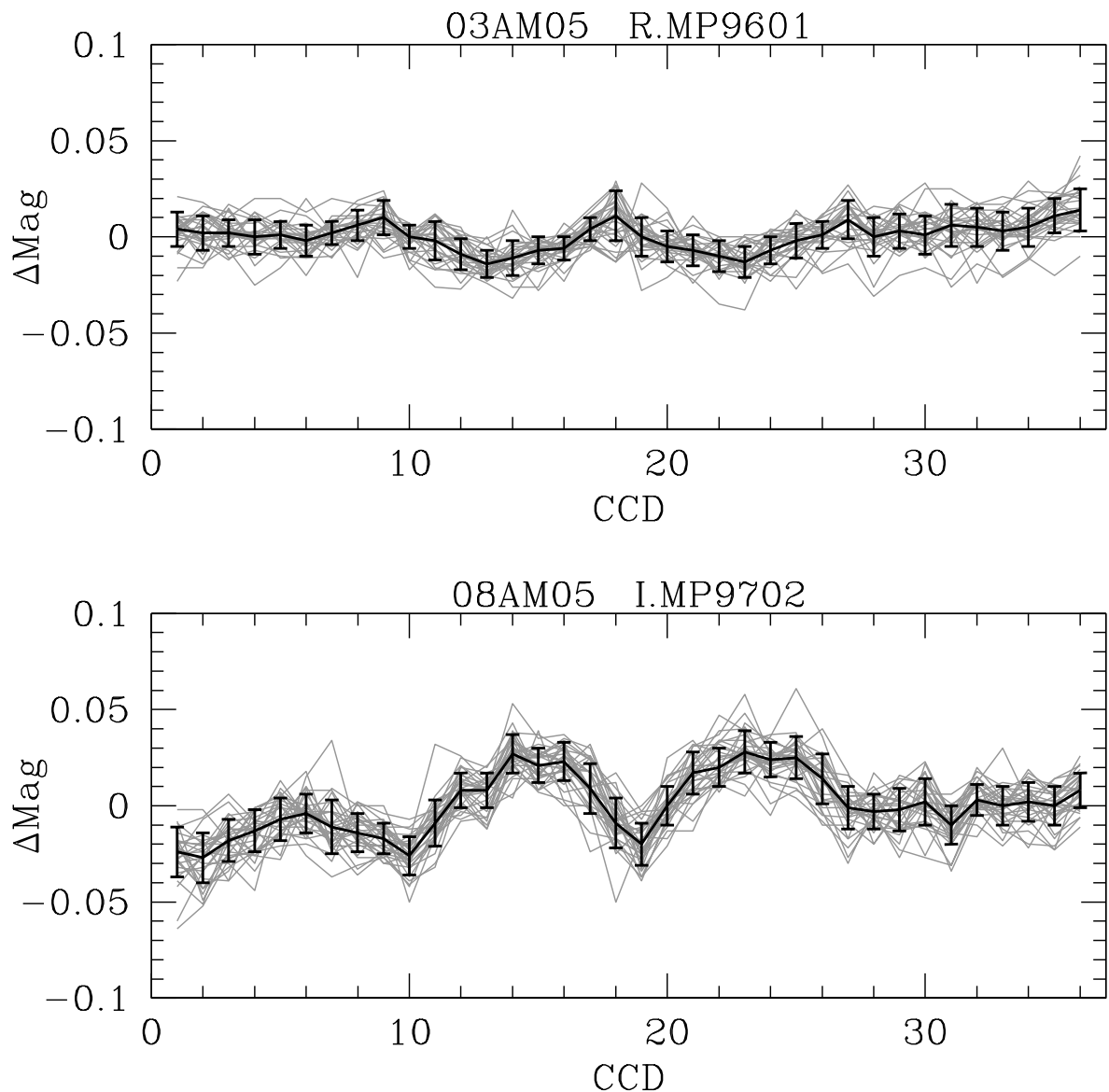}
\caption{ Differential zero-point variations.  The grey lines show the
individual zero-point offsets (SDSS-Elixir) as a function of CCD
number. The superimposed black line shows the median zero-point
offsets. There is clearly some scatter about the median, indicated by
the grey lines and by the error bars (1 $\sigma$).
The two panels are labelled by camera run ID and filter name.
The scatter in the top panel is 0.01 mags with a peak-to-peak
value of 0.03 mags.
The scatter in the top panel is 0.03 mags with a peak-to-peak
value of 0.08 mags.
}
\label{fig:dphotbw}
\end{figure}

The resulting differential zero-point measurements were aggregated by
filter and camera run ID (=CRUNID). For each of the CCDs, a median
photometric zero-point offset was computed for each filter/CRUNID
combination. The individual and median zero-point offsets were
plotted.  Two examples are shown in Figure \ref{fig:dphotbw}.  In
these plots the grey lines show the individual zero-point offsets
(SDSS-Elixir) as a function of CCD number. The superimposed black line
shows the median zero-point offsets. There is some scatter about the
median, indicated by the grey lines and by the error bars (1 sigma).
The pattern of offsets by CCD number reflects the mostly radial
pattern noted by \cite{regnault}.  The amplitude of the pattern varies
with filter, but also over time.  There is considerable variation in
the differential zero-point offsets.  For some runs the differential
offsets are quite small, as shown by the upper panel of Figure
\ref{fig:dphotbw}. For others it is substantial, (0.08 mags or
greater), as shown in the lower panel.  Going through the plots
sequentially, one sees definite changes over time. Further, at least
in some cases, the amplitude of the offsets is large enough to warrant
correction.

Zero-point corrections were determined for every CRUNID and filter
combination. These corrections were then applied to every CFHTLS
image which was not calibrated using the SDSS.

The next step was to measure the offsets of the MegaPipe mosaic as a
whole with respect to the SDSS.  To measure this for a given image,
the Elixir zero-point determined from the header keywords was
compared to the zero-point determined by comparing the instrumental
magnitudes of stars in the images to the magnitudes in the SDSS.
As before, the MegaCam archive was searched for all suitable
images. In this case, the criteria were slightly different. In order
for the Elixir zero-point to be valid, the night must have been
photometric. 

SkyProbe \citep{skyprobe} plots were used to determine whether or not
a given night was photometric. SkyProbe is a small camera mounted on
CFHT aligned with the telescope. It covers a 5~$\times$~7 field of
view down 12$^{th}$ magnitude. It monitors the flux from several
hundred stars over the course of the night.  Drops in the fluxes
indicate increase in atmospheric attenuation. Large variations in the
atmospheric attenuation indicate that the night was not photometric.
Plots for each night are available online\footnote{http://www.cfht.hawaii.edu/Instruments/Skyprobe/}.
 
The patches of bad SDSS photometry are small relative to the size of
the MegaPrime field of view so the requirement that all the pointings
must be different was relaxed. The new criteria were then:

\begin{itemize}
\item Each image must have an exposure time greater than 60 secs
\item Each image must lie within the SDSS
\item The night must be photometric. 
\end{itemize}

For each image, the average difference between the
SDSS and Elixir zero-point  over all the CCDs was measured. SExtractor was run on the
image with {\tt MAG{\underline\ }ZEROPOINT=0}. The resulting catalog
was astrometrically calibrated using the standard MegaPipe system. The
relevant part of the SDSS catalog was retrieved from the web. The
SDSS magnitudes were converted to MegaCam magnitudes using the
transformations described earlier.

The image catalog was matched to the SDSS catalog. The difference
between the instrumental magnitudes and the SDSS magnitudes
(transformed into the MegaCam filter system using Equation
\ref{eqn:trans}) yields the SDSS photometric zero-point. The Elixir
zero-point ${\tt ZP_{Elixir}}$ was calculated from the MegaCam image
header keywords as follows:
\begin{equation}
{\tt ZP_{Elixir}} = {\tt PHOT{\underline\ }C} + {\tt PHOT{\underline\ }K} \times ({\tt AIRMASS}-1) + 2.5 \times log_{10}({\tt EXPTIME})
\end{equation}
where 
{\tt PHOT{\underline\ }C} is the nightly instrumental zero point determined by Elixir,
{\tt PHOT{\underline\ }K} is the airmass coefficient,
{\tt AIRMASS} denotes the airmass, and
{\tt EXPTIME} is the exposure time.

The differential offset described above were also applied to the
Elixir zero-points. Note that because differential offsets average to
0 over the whole mosaic (by design), applying these offsets only
reduces the scatter of the individual CCD zero-points about the
average zero-point, but does not change the value of the average
zero-point.

The zero-point differences were then aggregated by filter and
CRUNID. The median zero-point difference was computed, as well as the
standard deviation about this median.  For some filter/CRUNID
combinations, no data was taken on photometric nights. For other
combinations, data was available for multiple photometric nights during
a single run. In a number of cases, there was variation in the
zero-point difference between two apparently photometric nights of the
same run. The zero-point differences might be stable within a single
night, but differences were found from night to night.
In addition,
there were occasionally variations in the zero-point differences within
a single night, indicating that the night was not photometric.

The zero-point differences for the filter/CRUNID combinations were
sorted into three categories, labeled ``No info'', ``Excellent'', and
``Marginal'' as follows:
\begin{itemize}
\item No info: No data was taken on photometric nights during this
  camera run with this filter.
\item Excellent: Data was taken on a least two photometric nights
  during this camera run, and at least 6 images were taken on those
  nights, and the scatter about the median zero-point difference is
  less than 0.02 magnitudes.
\item Marginal: There is some data, but it doesn't meet the criteria
  outlined above. Either data was not taken on more than one night, or
  less than 6 images were taken, or the scatter is large.
\end{itemize}

\begin{figure}
\plotone{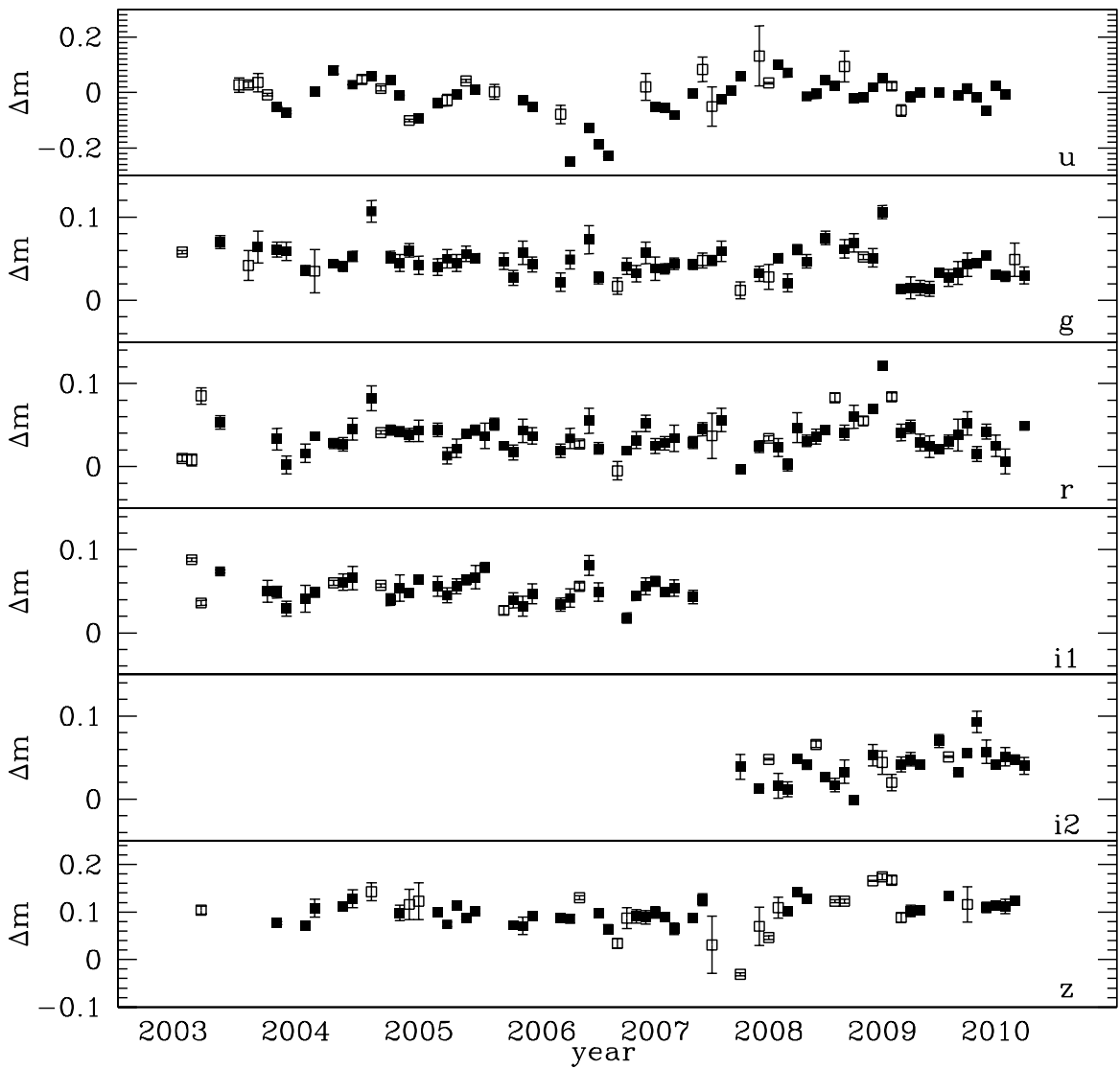}
\caption{Zero-point history for the MegaCam filters. The plot
shows the difference between the zero-points computed using the SDSS
as a reference to zero-points computed using the Elixir keywords.
The zero-point differences are plotted as a function of time.
$\Delta m = ZP_{SDSS} - ZP_{Elixir}$. 
Each point represents a single CRUNID.
The ``Excellent'' cases (as discussed in the text)
are shown as filled points. The ``Marginal'' cases are shown as open points . 
Note that the magnitude scales are not the same for all filters.
}
\label{fig:aphotbw}
\end{figure}

Figure \ref{fig:aphotbw} shows the zero-point
history of MegaCam for the \U\G\R\I\Z\ filters. 
Each point represents a single CRUNID.
The ``Excellent'' cases are shown as filled points
The ``Marginal'' case are shown as open points .  The ``No info'' cases
are not shown. The Elixir zero-point is not always
consistent with the SDSS zero-point. The amplitude of the variation
is typically about 0.05 magnitudes, except for the \U-band, for
which the amplitude is greater than 0.2 magnitudes.

Thus, two sets of corrections were applied to the Elixir zero-points: A
differential correction was applied to each CCD within the mosaic and a
global correction was applied the mosaic as a whole. Both sets of
corrections were applied as a function of CRUNID.  Applying these
corrections greatly increased the consistency of the photometry, as
discussed in Section \ref{ssec:qphotom}

\subsection{Using Overlapping Images}
\label{ssec:photomNeigh}

Some of the CFHTLS data neither lie in the SDSS nor were taken on a
photometric night. These data can not be photometrically calibrated by
either of the preceding methods. However, if such an image overlaps another
which can be calibrated by one the preceding methods, it in turn
can be calibrated.

For the Deep fields, this was fairly straightforward. Since all the
images in a field lie at same position, in-field photometric standards
using images taken on photometric nights were established.  Those
standards were then used to calibrate all the data for that field.  In
fact, the self-consistency of the photometry was used to determine
which nights were photometric.  The two Deep fields which do not lie
in the SDSS, the D1 and the D4, were calibrated in this manner.  The
initial photometry was computed for each image using the Elixir
zero-points. Catalogs were generated for each image based on these
zero-points and the catalogs were cross-referenced to each other to
determine image-to-image variations in photometry.  The photometric
consistency was checked over each night. Any night showing large
variation in photometry from image to image was deemed
non-photometric. (Interestingly, some of these nights appear
completely photometric based on the SkyProbe measurements.)  While the
presence of large variation indicates that the night was not
photometric, the absence of such variations does not guarantee that
the night was photometric.  The photometric catalogs on the apparently
photometric nights were compared over the run of the survey. Again, if
the photometry for a given night was not consistent with that of other
nights, it was flagged as non-photometric. In the end, a minimum of 5
nights per filter and per field were identified as both photometric
and consistent. The images from these nights were stacked and a
catalog generated from the resulting image.  This catalog became the
photometric reference for that field and filter.

A similar method was used for the Wide fields.  The pointings within a
field overlap at the edges, allowing photometric comparisons.  Each
filter was processed separately.  First, the photometry was
homogenized within each pointing.  In the simplest case, all the
images within a pointing could be directly calibrated using either the
(corrected) Elixir zero-points or the SDSS. These pointings were
flagged as probably photometric (PP).  If only some of the images of a
pointing could be directly calibrated, these images were used as
reference for the others. These pointings were also flagged as
PP.  If none of the images could be calibrated
directly, one of the images was used as a photometric reference for
the others. The photometry of the images in this pointing would be
self-consistent within itself, but probably not with adjacent
pointings. These pointings were flagged as not photometric (NP).

Next, the photometric consistency between pointings was
checked. Consistency here means a systematic zero-point difference of
less than 0.03 magnitudes.  If a pointing previously flagged PP was
found to be inconsistent with any other PP or DP pointing, then its
status was downgraded to NP. On the other hand, if a pointing was
consistent with at least two other adjacent pointings its status was
upgraded to definitely photometric (DP).

Having identified the DP pointings, the next step was to calibrate
adjacent pointings. There are typically a few hundred stars in the
overlap region between two adjacent pointings. Only pointings which
overlap along an edge were calibrated in this manner, not pointings
which only overlap at the corners.  The random error associated with
transferring the zero-point in this manner is typically 0.05 mags per
star. When 300 stars are used, the random error drops to below 0.002
mags.  Once a new pointing was calibrated by overlap, it was checked
for consistency with previously calibrated pointings. If it was
consistent, it was flagged as DP and used to calibrate other adjacent
pointings.  Eventually all pointings in a field were calibrated. In
principle transferring the zero-point in this manner could cause an
increase in photometric zero-point error as the number of transfers
increases.  In practice all pointings were at most 2 steps away from a
DP pointing. The W1, W2 and W4 fields were calibrated in this manner.

\section{IMAGE STACKING}
\label{sec:stack}

The calibrated images were resampled and coadded using the program
\swarp\ \citep{swarp}. \swarp\ removed the sky background from each
image so that its sky level was 0. It scaled each image according to
the photometric calibration described in Section \ref{sec:photom}.
\swarp\ then resampled the pixels of each input image to remove the
geometric distortion measured in Section \ref{sec:astrom} and placed
them in the output pixel grid, which is an undistorted tangent plane
projection.  A ``Lanczos-3'' interpolation kernel was used as
discussed in Section 5.6.1 of Bertin (2004).  The values of the
flux-scaled, resampled pixels for each image were then combined into
the output image by taking a median. A median is noisier than an
average, but rejects image defects such as cosmic rays.  The loss in
depth is small (about 0.1 magnitudes). The optimum would be some
sort of artificial skepticism method \citep{artskep}, but this
is not yet an option in \swarp.

The input images were weighted with mask images provided as part of the
Elixir processing. The mask images had the value 1 for good data and
0 for pixels with known image defects. An inverse variance weight map
was produced along with each output image.  This could be used as an
input when running SExtractor on the stack.

The resulting stacks measure about 20000 pixels by 20000 pixels or
about 1 degree by 1 degree and are about 1.7 Gb in size. They have a
sky level of 0 counts. They are scaled to have a photometric
zero-point of 30.000 in AB magnitudes - that is to say, for each
source:
\begin{equation}
AB_{magnitude} = -2.5 \times \log_{10}({\rm counts~in~ADU}) + 30.000
\end{equation}

\section{CATALOG GENERATION}
\label{sec:cat}
\subsection{Single Image Catalogs}
\label{ssec:simplecat}

SExtractor was run on each image to produce a catalog.  The images
were filtered before detection.  The detection criteria were
that 3 adjacent pixels had to have flux levels 1 $\sigma$ above the
background.  The SExtractor's deblend contrast parameter was set to
0.002.  Table \ref{tab:sex} shows the relevant SExtractor detection
parameters.  The inverse variance weight maps described in the
previous section were used.  The resulting catalogs, one per pointing
and filter, are available on the web as described in Section
\ref{sec:distro}.

\subsection{Catalog Masking}
\label{ssec:catmask}
The individual catalogs were masked to identify areas where the
detection and photometric measurements of sources may
compromised. These areas include areas around bright stars,
diffraction and bleed spikes from the brightest stars and
satellite/meteor trails. Also, in some cases the dither pattern of the
input images was insufficient to provide uniform depth across the
pointing. An automatic detection method was used to find the bright
stars and the diffraction/bleed spikes. This was supplemented with
laborious hand masking. 

\begin{figure}
\plotone{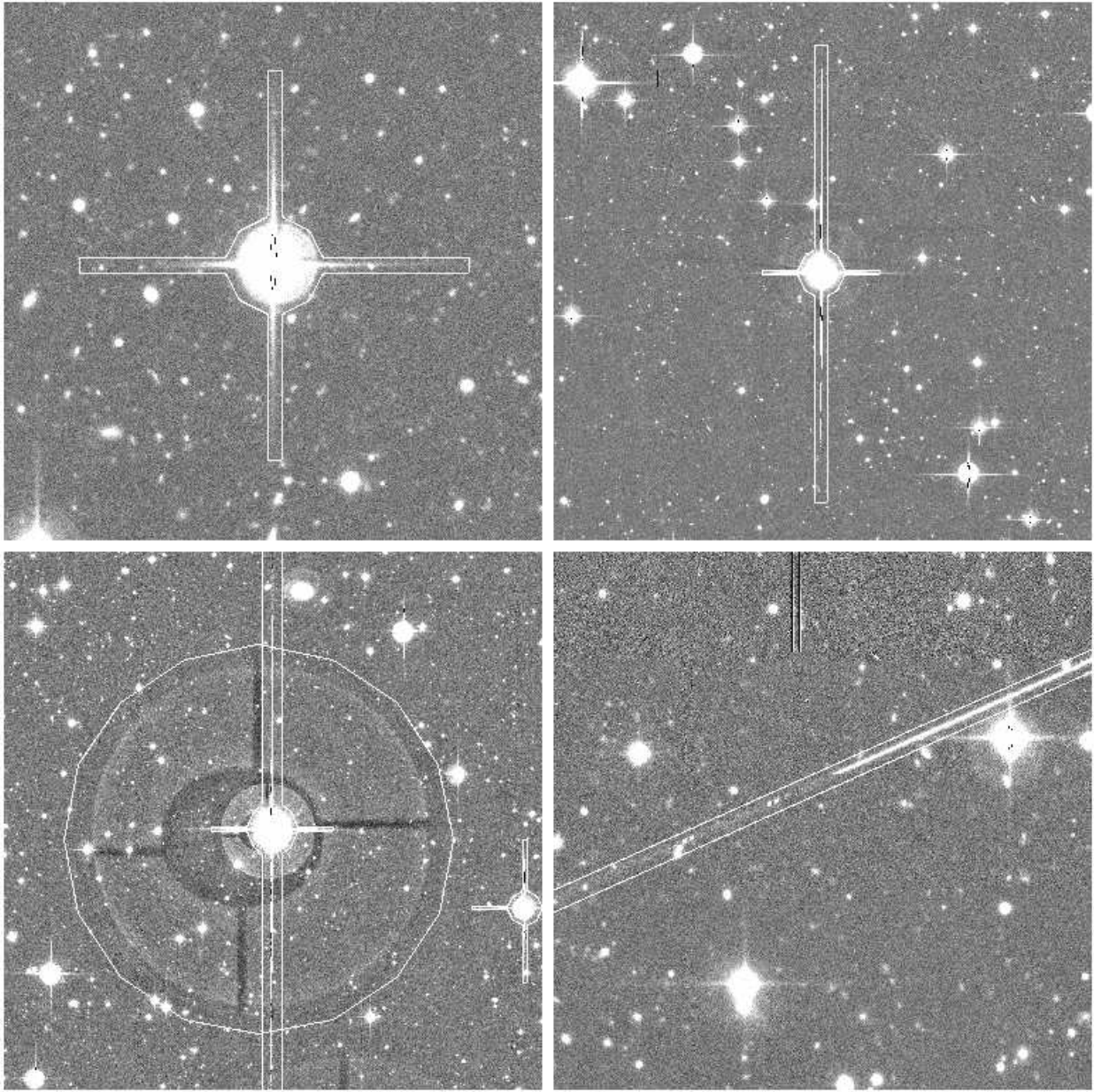}
\caption{ Examples of masking.  The top left panel shows bright star
  masking, including masking of the diffraction spikes.  The top right
  panel shows the same, but this time the mask also covers the bleed
  trails from the star in the y-direction.  The bottom left panel
  shows pupil image masking; the figure shows a circular mask, which
  is slightly offset from the center of the star.  The bottom right
  panel shows meteor trail masking.  Note that the different
  panels are not at the same pixel scale.  }
\label{fig:maskallbw}
\end{figure}

The areas around bright stars were masked. The Guide Star Catalog 2
\citep{gsc2} was queried for the position and magnitude of stars
brighter than $15^{th}$ magnitude. The stars were masked with a
circular mask covering the core of the star and a cross-shaped area
covering the diffraction spikes.  The size of the mask scaled with the
brightness of the star.  The top left panel of Figure
\ref{fig:maskallbw} shows an example.  Stars brighter $13^{th}$
magnitude often produce bleed trails in the images extending along the
y-axis of the CCDs. The masking algorithm followed the bleed trails
until they blended into the sky background, extending the cross part of
the stellar masks in the vertical direction. The top right panel of Figure
\ref{fig:maskallbw} shows an example.

Stars brighter than $9^{th}$ magnitude also produce a visible pupil
image.  These halos have a radius of about 1100 MegaCam pixels
(approximately 3.5 arcminutes). The center of the halos are offset
from the star that created it towards the center of the MegaCam field;
the size of the offset is 0.022 times the distance of the star to the
center of the field, as noted in Section 3.4 of \cite{CARS}.  The
bottom left panel of Figure \ref{fig:maskallbw} shows an example of a
pupil mask.

In addition, meteor trail residuals were masked. Since meteor trails
only show up in one image, and since a median is used to combine
the images, these are mostly removed during the stacking process,
leaving at most some residual noise.  However, in some places (the
gaps between CCDs of the MegaCam mosaic and over bad columns) less
than the full number of input images was available, and the meteor
trails were visible in the output image. The trail masking method
described in Section 3.4, item 3. of \cite{CARS} was
attempted. However as noted in that section, it is not 100\% reliable.
Additional hand masking of meteor trails was required.
The bottom right panel of 
Figure \ref{fig:maskallbw} shows an example of meteor trail masking
near the edge of a mosaic.

Masks were generated for each image of the survey. The masks were
checked by eye and minor modifications made.  The masks are available
for download as ds9 region files\footnote {\tt
  http://www.cadc-ccda.hia-iha.nrc-cnrc.gc.ca/community/CFHTLS-SG/docs/catdoc.html\#mask}.

\subsection{Merged Catalogs}
\label{ssec:mergecat}

To increase the user-friendliness of the catalogs, the individual
catalogs were merged. Each merged catalog contains measurements in
all 5 of the \U\G\R\I\Z\ MegaCam filters and covers an entire survey (either
the Wide or the Deep).

Separate catalogs were generated using each of the 5 bands as a
reference/detection image. Thus there is an \U-selected, \G-selected,
\R-selected, \I-selected and \Z-selected catalog for each survey each
suited to different science goals. (e.g. \I-selected for general galaxy
population studies, \Z-selected if you are looking for \I-band
dropouts, \G-band for galactic stars, etc.). Although the catalogs are
selected in a single filter, measurements are made in all 5 filters
for each catalog, removing the need to consult multiple
catalogs. There are thus a total of $5\times 2 =10$ separate catalogs,
each complete in itself, but with different detection characteristics.

The procedure to produce the merged catalogs is:
\begin{itemize}
\item Individual catalog generation: SExtractor is run in double-image
  mode on the individual images.
\item Catalog masking: Areas where photometric measurements and
  detections may be compromised are masked as described in section \ref{ssec:catmask}.
\item Catalog merging: The masked catalogs for each pointing are
  combined and redundant sources (ones which appear in more than one
  individual catalogs) are removed.
\end{itemize}

Before settling on this procedure, A number of other methods were
considered and ultimately rejected:

Instead of generating catalogs pointing-by-pointing and then merging
the catalogs, one could SWarp together all the images for each Wide
field into a single enormous image and run SExtractor on that
image. This has the advantage of eliminating the catalog merging step,
which involves a fair amount of book-keeping. The disadvantage is that
these images are unwieldy. They would measure up to 140 000 pixels
across and be 100 Gb in size. While it may be possible (although
difficult) to run SExtractor on such images, one cannot distribute
such images over the web. Users would be presented with catalogs for
which they cannot view the source image.

Instead of generating separate catalog for each filter, one could build
some sort of master image containing flux from each band, for example
a $\chi^2$ image. One could then use SExtractor in double-image
mode with the master image as a detection image. This has the
advantage of simplicity: it produces a single catalog. However, the
5 bands are not even in depth. The z-band in particular is much
noisier than the other 4. While adding the bands together produces a
deeper image in theory (you now have photons from all wavelengths), in
practice it is also noisier. Leaving out the noisier bands (\U\Z) means
that the resulting catalog is no longer suitable for all purposes.

Instead of publishing a separate catalog for each filter, one could
merge the 5 catalogs into one. One could cross-identify sources
common to different catalogs and merge their entries into a single
entry. This again has the advantage of producing a single catalog
rather than several, and (if done correctly) eliminates the
disadvantages of the previous master-image method. The trick, of
course, is accurately and reliably cross-identifying the sources
between filters. For bright, well separated sources this is easy. The
MegaPipe images are registered to very high accuracy, making
cross-identification by position fairly accurate. However the image
quality and depth varies between different filters; objects that
appear as single source in one filter may appear as two sources in
another filter. Even if the cross-identification can be done
correctly, the bookkeeping is non-trivial. That being said, it has
work successfully for the SDSS and this option may be explored in the
future.

In short, the easiest way to make a clean catalog is to: 
\begin{itemize}
\item merge the different pointings at the catalog level rather than at the image level 
\item not to merge the catalogs for different filters but to generate multiple catalogs, one per filter. 
\end{itemize}

\subsubsection{Individual Catalog Generation}

SExtractor was run on the 5 images from each pointing using
``double-image mode''. In this mode, the detection of objects is done
in one image (the reference image) and photometric (and other)
measurements are done in the other (the measurement image). All
possible image combinations were run for a total of 25 catalogs per
pointing (5 possible detection images $\times$ 5 possible measurement
images).

\subsubsection{Catalog merging}

The final step was to merge the catalogs. For the Deep fields, which
do not overlap, the catalogs for the individual fields were simply
concatenated. The pointings of the Wide fields however, overlap
slightly. Simply concatenating the individual catalogs would lead to
many sources being double-counted.

Boundaries were determined at the edges of each Wide pointings.  These
boundaries lie at the point where the effective exposure time of the
images (as indicated by weight map) drops to half of the nominal
value. Where the boundaries overlapped with adjacent pointings, the
boundaries were shrunk until the pointings only overlapped by
$\theta_{overlap}=20$ arcseconds. This size is enough to accommodate
the positional uncertainty of even fairly extended objects.  Inside
the overlap regions, objects from different pointing catalogs lying
with $\theta_{match}=0.5''$ of each other were deemed to be the same
object; the second entry was removed.  Outside of the overlap regions,
all sources from the individual catalogs were included in the merged
catalog.

There are two parameters in this method: $\theta_{overlap}$ and
$\theta_{match}$. Changing either parameter significantly (doubling or
halving it) had only a very small effect on the final merged catalogs:
the total number of sources would change by a few hundred in a catalog
of 30 million sources.

\section{QUALITY CONTROL}
\label{sec:qual}
\subsection{Astrometry}
\label{ssec:qastrom}

\subsubsection{Internal Astrometric Consistency}
\label{sssec:qastroin}

The internal accuracy was checked by running SExtractor on each stacked
image in every band and obtaining catalogs of object positions. The
positional catalogs for each band were matched to each other and
common objects identified. If the astrometry was perfect, then the
position of an object in each band would be identical. In practice,
there are astrometric residuals. Examining these residuals gives an
idea of the astrometric uncertainties.

\begin{figure}
\plotone{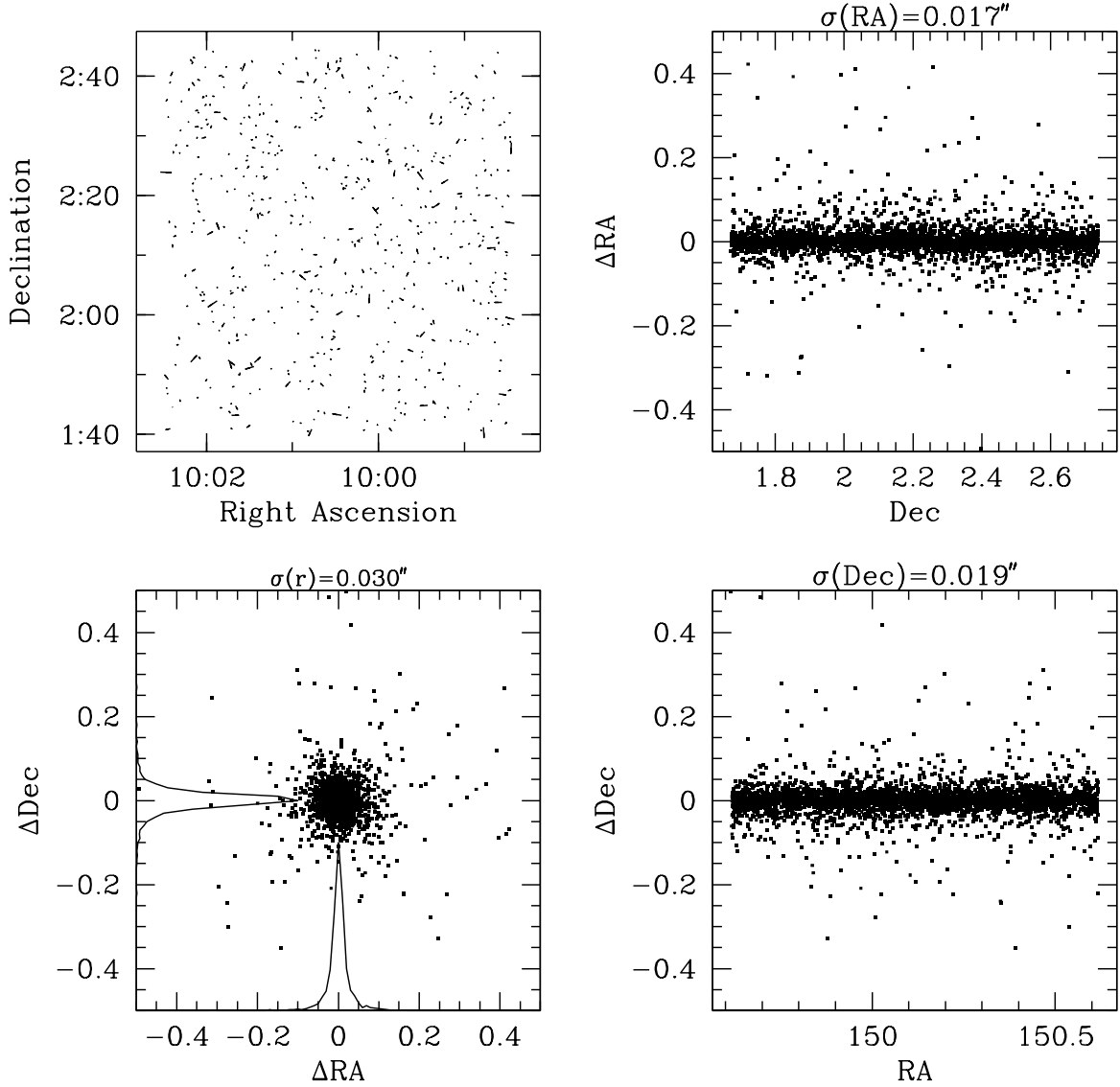}
\caption{An example of internal astrometric residuals.  The top left
  panel shows the direction and size (greatly enlarged) of the
  astrometric residuals as line segments.  The bottom left panel shows
  the astrometric residuals in RA and Dec. The histograms show the
  relative distribution of the residuals in both directions.  The two
  right panels show the residuals in RA and Dec as functions of Dec
  and RA respectively.  }
\label{fig:cfhtls.astint.bw}
\end{figure}

Figure \ref{fig:cfhtls.astint.bw} shows checks on the internal astrometry
between the \G\ and \R\ band images for the D2 field.  The top left
quarter shows the direction and size (greatly enlarged) of the
astrometric residuals as line segments. This plot is an important
diagnostic of astrometry because, while the residuals are typically
quite small, there are outliers in any distribution. As long as these
outliers are relatively isolated from each other and pointing in
random directions all is well. Conversely, if there are a number large of
residuals in close proximity to each other, all pointing the same
direction, this indicates a systematic misalignment between the two
images in question. The figure shows no such misalignments.  Similar
plots were made for every possible pair of images in the CFHTLS.
No misalignments or other problems were found, indicating
there are no systematic astrometric errors.

The two right panels show the residuals in RA and Dec separately,
which are about 0.02 arcseconds (68\%-tile).  The bottom left
quarter shows the astrometric residuals in RA and Dec. The 
histograms show the relative distribution of the residuals in both
directions. The 68\%-tile of the residuals is 0.03 arcseconds
radially, about $\sqrt{2}$ larger than the residuals in any one
direction.

In general, the internal astrometric uncertainties are slightly larger
than the example shown in Figure \ref{fig:cfhtls.astint.bw}.  Similar plots were
made for every possible pair of images in the CFHTLS.  The typical
internal astrometric residuals are 0.04$''$\ in any one
direction. However, no misalignments or other problems were found,
indicating there are no systematic astrometric errors.

\subsubsection{External Astrometric Accuracy}
\label{sssec:qastroex}

\begin{figure}
\plotone{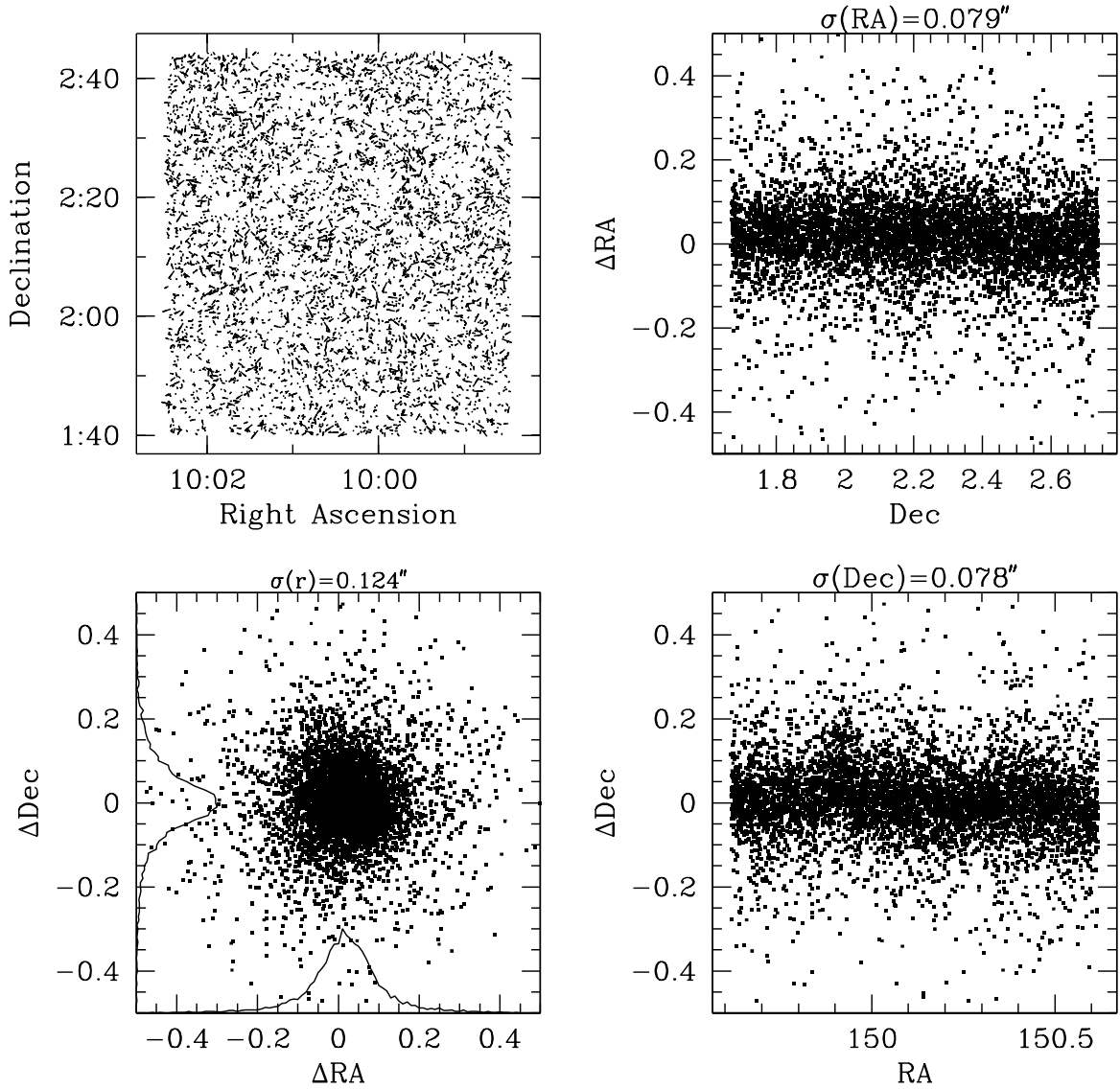}
\caption{An example of external astrometric residuals with respect to the SDSS.
The panels have the same meaning
as in Figure \ref{fig:cfhtls.astint.bw}.}
\label{fig:cfhtls.astext.bw}
\end{figure}

Accurate astrometry relative to external systems is relevant for
follow-up observations with other telescopes.  External accuracy was
checked by matching each catalog for each field and filter back to the
astrometric reference catalog. Again, the scatter in the astrometric
residuals is a measure of the uncertainty and the presence of any
localized large residuals indicates a systematic shift.  Figure
\ref{fig:cfhtls.astext.bw} shows an example of the astrometric residuals of one of
the pointings with respect to the SDSS. The panels have the same
meanings as in Figure \ref{fig:cfhtls.astint.bw}.  The external residuals are larger than
the internal residuals.  Similar plots were generated for all
pointings and bands.  On average, the astrometric residuals are 0.07
arcseconds with respect to the SDSS and 0.18 arcseconds with respect
to NOMAD. No systematic offsets are visible, indicating that they are
smaller than the random residuals, if indeed they exist.

\subsection{Photometry}
\label{ssec:qphotom}

\subsubsection{Internal Consistency}
\label{sssec:qphotin}

\begin{figure}
\plotone{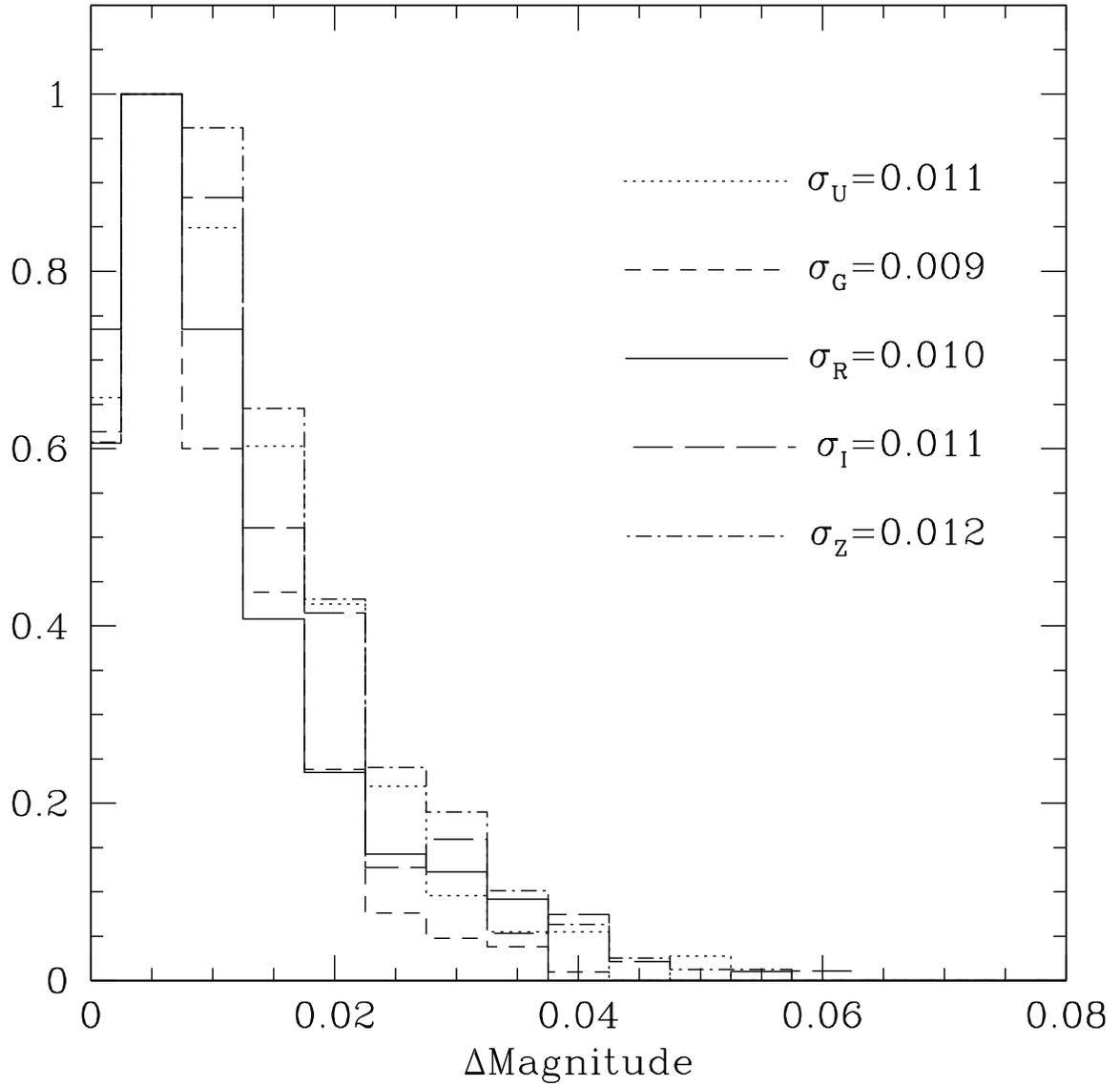}
\label{fig:widenphotbw}
\caption{Internal photometric checks for the Wide Fields.  The
  histograms show the photometric zero-point offsets between adjacent
  fields. The histograms are split by filter as indicated by the different line types.  }
\end{figure}

The edges of the Wide pointings overlap. Comparing the photometry of
objects in the overlap regions gives a measure of the photometric
consistency.  Systematic difference in photometry indicate zero-point
offsets.  Figure \ref{fig:widenphotbw} shows a histogram of the
pointing-to-pointing zero-point differences. In comparing two
pointings, the sign of the zero-point difference is irrelevant, so the
absolute value of the differences is plotted here. The average
offsets are about 0.01 magnitudes, with a tail out to 0.05
magnitudes.

\subsubsection{External Accuracy }
\label{sssec:qphotex}

\begin{figure}
\plotone{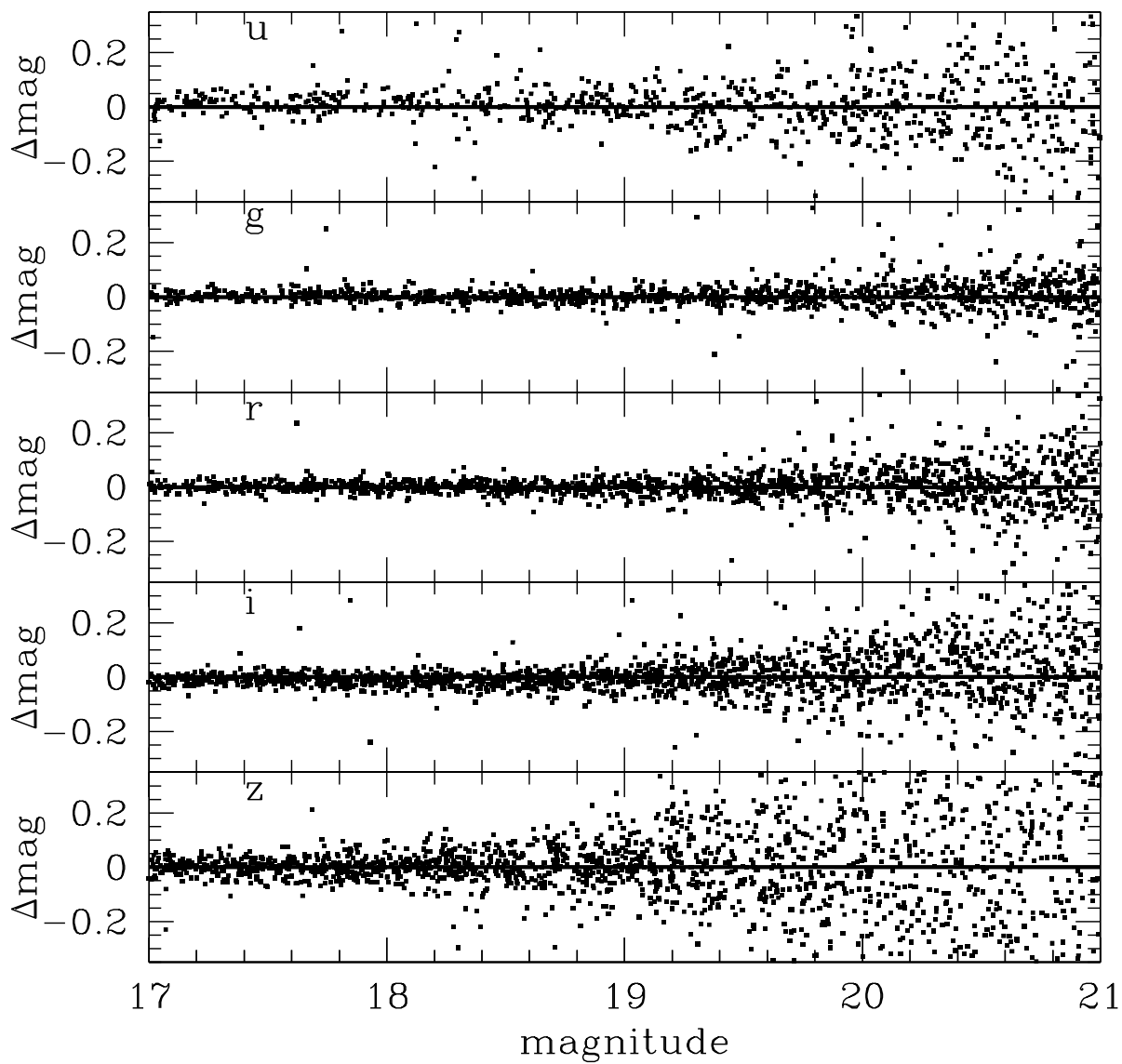}
\caption{An example of a comparison of the photometry between the
  CFHTLS and the SDSS. The SDSS magnitudes have been transformed to
  the CFHTLS system.  At bright magnitudes, for \I-band,
  there is a deviation, caused by the brightest stars
  saturating. In \U-band the scatter is larger because of the larger
  difference between SDSS-$u'$ and MegaCam-\U, but the center of the
  distribution remains close to 0.
}
\label{fig:magcompubbw}
\end{figure}

At least some of the pointings of each of the Wide fields as well as
the D2 and D3 fields lie within the SDSS. This makes it possible to
directly compare the magnitudes in those fields to an external
reference. Figure \ref{fig:magcompubbw} shows a typical comparison
between the SDSS (transformed to the MegaCam system as described by Equation
\ref{eqn:trans})
and the CFHTLS for the 5 bands. The agreement is very good. At bright
magnitudes, for g, r and i bands, there are deviations caused by stars
saturating in the MegaCam images. There is no evidence for systematic
shifts greater than 0.01 magnitudes. There is also relatively little
scatter (at least at moderate magnitudes), which argues that the color
terms in the SDSS-MegaCam transformation are fairly accurate. The
\U-band shows greater scatter than the other bands; this is caused by the
larger and more uncertain color terms in the transformation between
the SDSS and MegaCam photometric systems. 

\begin{figure}
\plotone{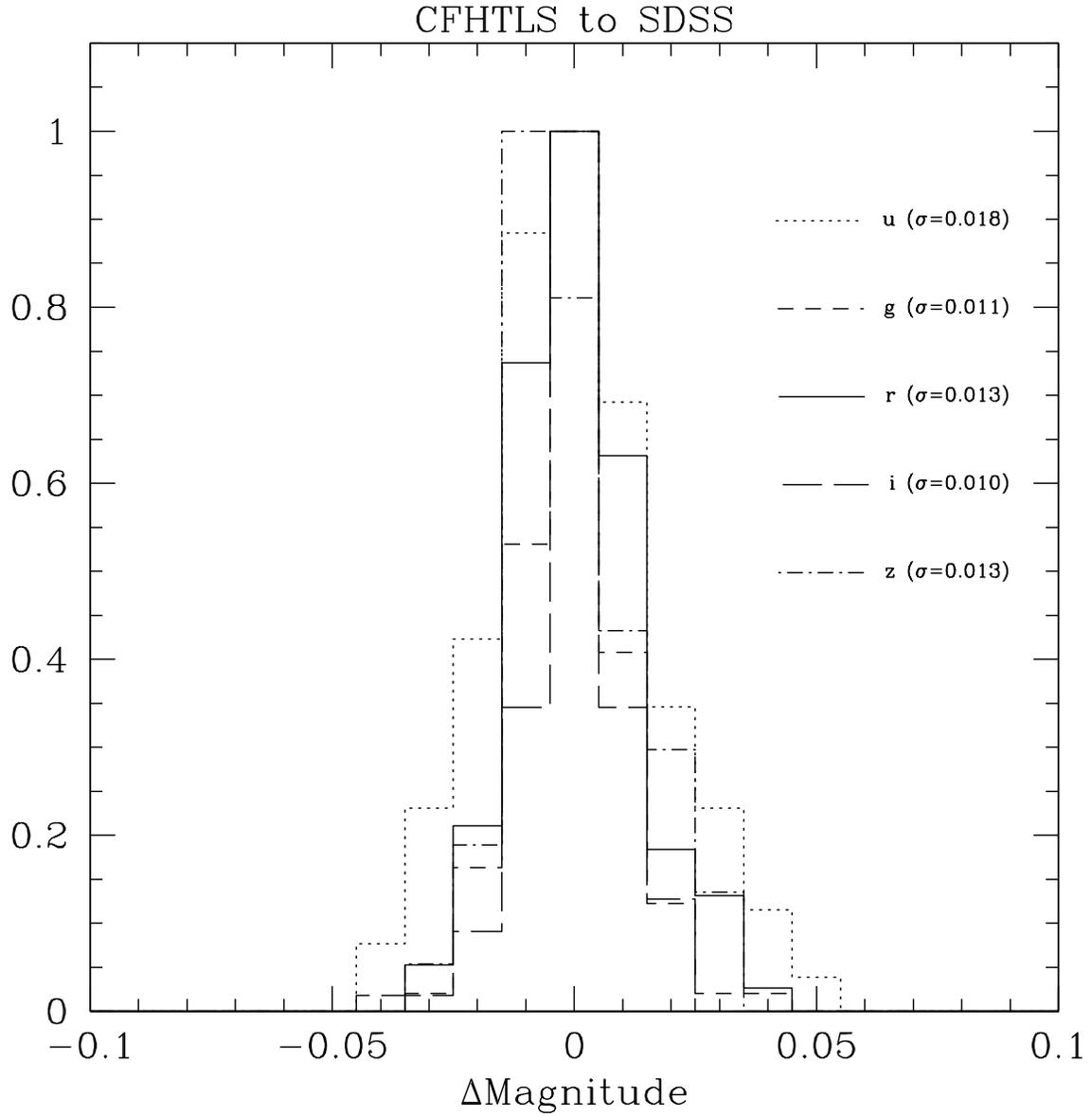}
\caption{External photometric checks for the Wide Fields. The
  histograms show the photometric zero-point offsets between the
  CFHTLS and the SDSS. The histograms are split by filter as in the
  previous figure. The offsets are typically slightly larger than the
  internal photometric residuals,.}
\label{fig:wide.sdss.phot.bw}
\end{figure}

Similar comparisons have been done for every pointing overlapping the
SDSS. Figure \ref{fig:wide.sdss.phot.bw} summarizes the results. The
offsets are typically slightly larger than the internal photometric
residuals shown above, ranging from 0.010 magnitudes in \I\ to 0.018
for the \U-band.

This test is not a completely independent test of photometry because
in many cases the photometric zero-point was derived from the SDSS.
However, for many pointings it is independent: for example, as shown
in Figure \ref{fig:W4.bw.lay} the ``+0'' row of the W4 field lies only
partially within the SDSS. These pointings were not directly
calibrated using the SDSS, but rely on the Elixir calibration and
adjacent pointings for their photometric zero-points. However when the
CFHTLS magnitudes of objects in the overlap region of these pointings
are compared with SDSS magnitudes, they agree to within the errors
discussed above.

As an additional test, the W3 field (which lies entirely within the
SDSS) was reduced a second time, without any reference to the SDSS.
The reduction was done using only the modified Elixir photometric
calibration as described in sections \ref{ssec:photomElixir} and
\ref{ssec:photomNeigh}.  The zero-points thus derived were found to be
consistent with the zero-points derived from the SDSS to with 0.01
mags RMS, 

\citet{regnault} did an extremely thorough analysis of the photometric
calibration of the SNLS. This analysis is completely independent of
the current analysis, and ties its photometric zero-points to
\citet{Landolt2007} standards. They published tertiary standards in
for the 4 Deep fields. The present photometry was compared to these
tertiary standards. After correcting for the zero-point difference
(the Regnault standards are in a natural magnitude system, which is
approximately Vega-based, where as the MegaPipe photometry is in the AB
system), the zero-point offsets were found to be approximately 1\%.
Again, the offsets in the \U-band were found to be slightly larger,
approximately 2\%. The scatter about the offsets across the field

\subsubsection{Magnitude limits}
\label{sssec:maglim}

The limiting magnitudes of the images were determined by adding fake stars
and galaxies to the images and then trying to recover them using the
same parameters used to generate the real image catalogs.

The fake galaxies used were taken from the images themselves, rather
than adding completely artificial galaxies. A set of 40 bright,
isolated galaxies were selected out of the field and assembled into a
master list. Postage stamps of these galaxies were cut out of the
field. The galaxies were faded in both surface brightness and
magnitude through a combination of scaling the pixel values and
resampling the images.

To test the recovery rate at a given magnitude and surface brightness,
galaxy postage stamps were selected from the master list, faded as
described above to the magnitude and surface brightness in question
and then added to the image at random locations. SExtractor was then
run on new image. The fraction of fake galaxies found gives the
recovery rate at that magnitude and surface brightness. An
illustration of adding the galaxies is shown in Figure 11 of
\cite{gwyn2008}.

To test the false-positive rate, the original image was multiplied by
-1; the noise peaks became noise troughs and vice-versa. SExtractor
was run, using the same detection criteria. Since there are no real
negative galaxies, all the objects thus detected are spurious.

\begin{figure}
\plotone{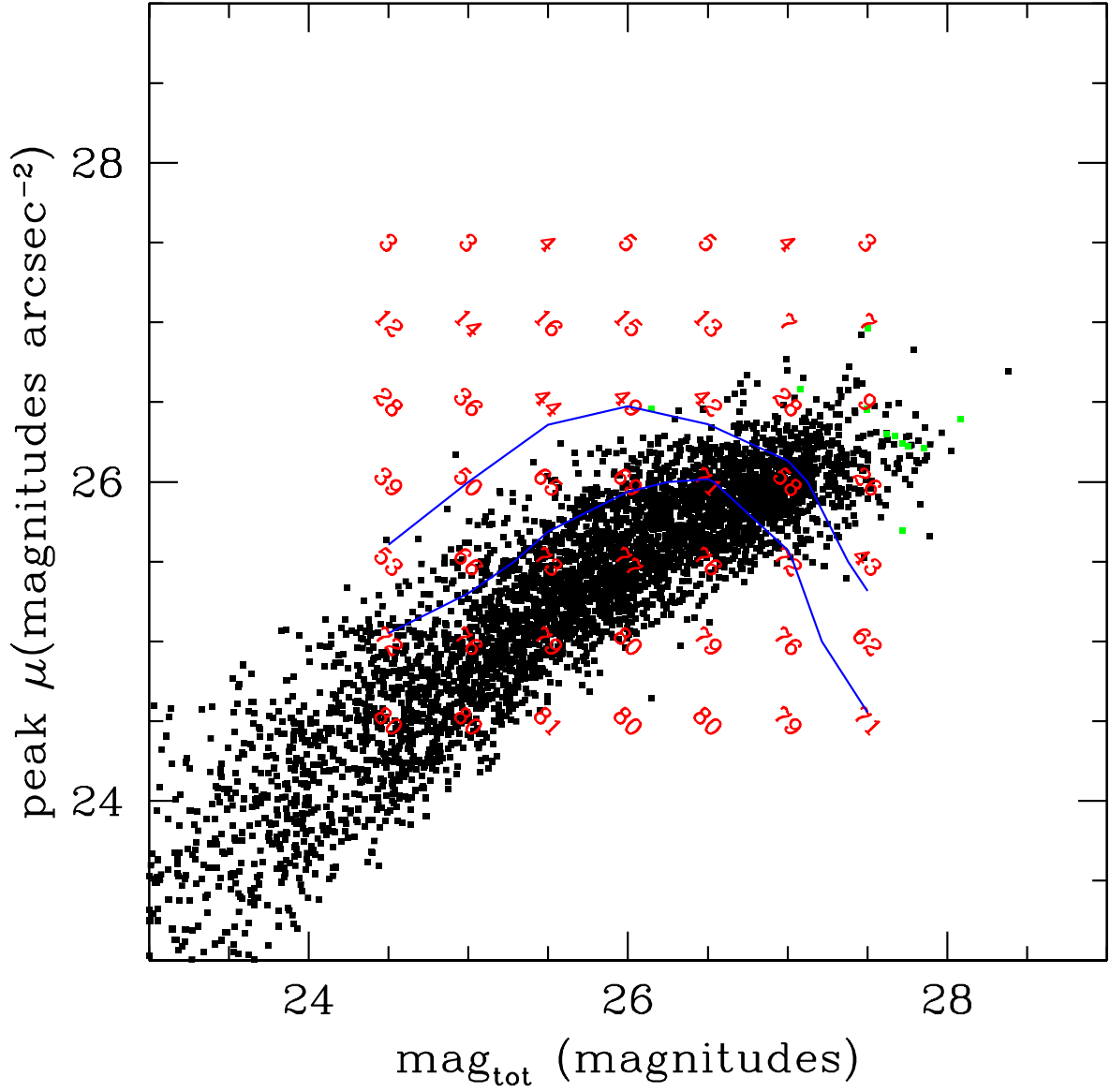}
\caption{ Completeness limits in magnitude and surface brightness.
  The black points are real objects.  The green points show the
  false-positive detections.  The red numbers show the percentage of
  artificial galaxies that were recovered at that magnitude/surface
  brightness.  The blue contour lines shows the 70\% and 50\%
  completeness levels.  }
\label{fig:maglimauto}
\end{figure}

The magnitude/surface brightness plot in Figure \ref{fig:maglimauto}
shows the results of such tests.  The black points are real
objects. The bottom edge of the black points is the locus of
point-like objects. The green points show the false-positive
detections, located at faint magnitudes and surface brightnesses.  The
red numbers show the percent of artificial galaxies that were
recovered at a given magnitude/surface brightness. Superimposed on the
numbers are two contour lines in blue, at 50\% and 70\% completeness.

Deriving a single limiting magnitude from such a plot is slightly
difficult. The cleaner cut in the false positives seems to be in
surface brightness. Extended objects become harder to detect at
brighter magnitudes whereas stellar objects are detectable a magnitude
or so fainter.

\begin{figure}
\plotone{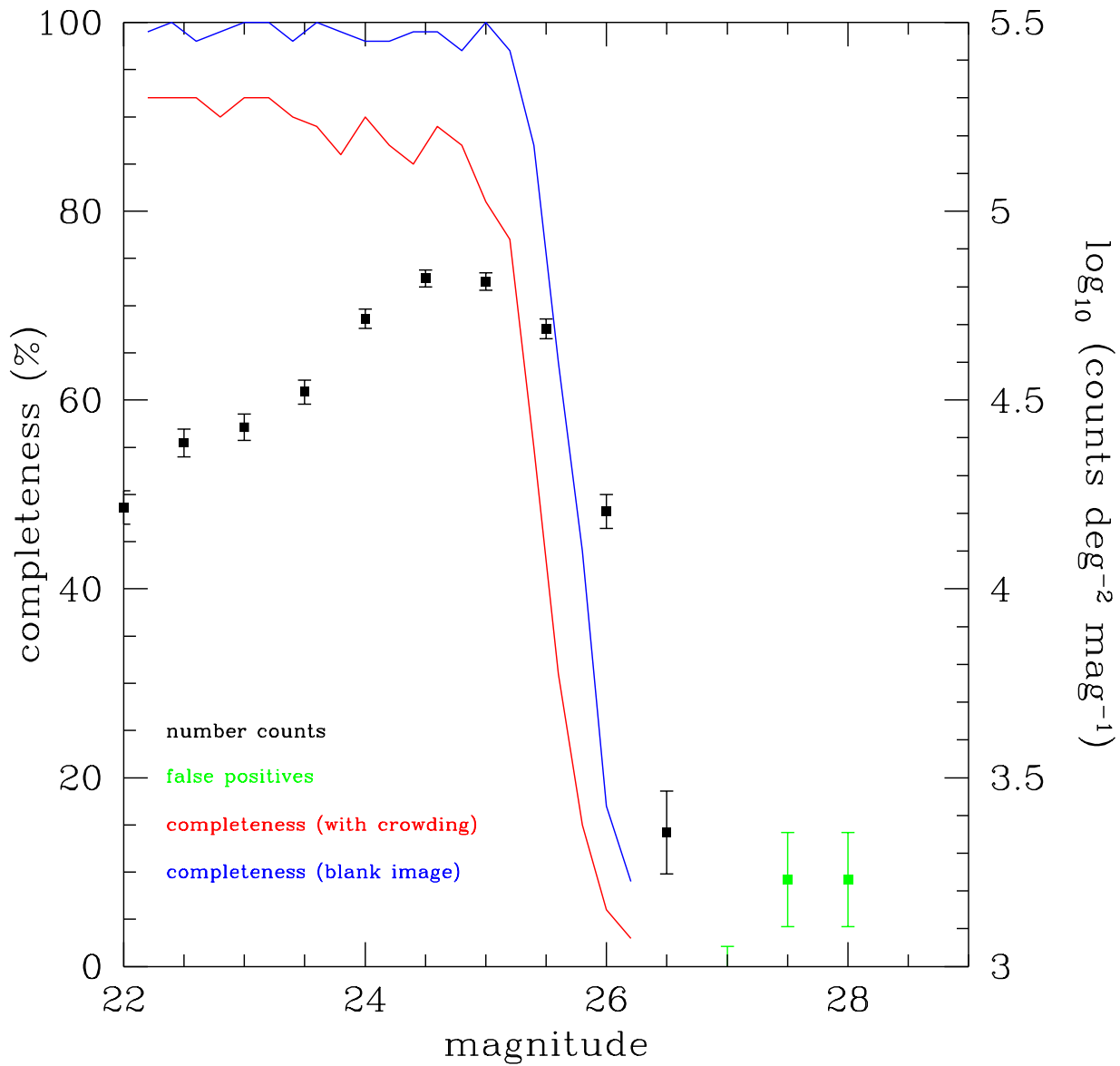}
\caption{ Point source completeness limits. The blue line shows the
  fraction of artificial point sources that can be recovered from the
  blank image. The red line shows the same for the original image.
  The black points show the number counts of real sources. The green
  points show the number counts for false positive detections.
  The completeness limits for this Wide pointing is 25.7 magnitudes.  }
\label{fig:magstar}
\end{figure}

Point source limiting magnitudes were also calculated. Point sources
were added to the images in a similar manner to the above, but only
scaled with magnitude. For the Deep fields in particular, the images
were effectively crowded. An artificial source added to the image
stood a significant chance of ending up close enough to a real source
that it would not be detected. To compensate for this, sources were
added to two images. The first was just the original image. The second
was the original image with all the real sources removed and their
pixels replaced with values matching the background noise
characteristics; {\it i.e.}, a blank image. The differences between
the two completeness limits is shown in Figure \ref{fig:magstar}.
This figure shows the completeness limits for a Wide pointing. The
blue line shows the fraction of artificial point sources that can be
recovered from the blank image. The red line shows the same for the
original image. It is consistently a few 10$^{ths}$ of a magnitude
less deep. The black points show the number counts of real
sources. The green points show the number counts for false positive
detections. The 50\% completeness limit for this image is 25.7
magnitudes. The 50\% blank-image completeness limits are the values
quoted as limiting magnitudes in Tables 1, 2 and 3.

\section{DISTRIBUTION}
\label{sec:distro}

The MegaPipe processed images are available
via webpages at the Canadian Astronomy Data Centre\footnote{
\tt http://www.cadc-ccda.hia-iha.nrc-cnrc.gc.ca/community/CFHTLS-SG/docs/cfhtls.html}.
There are a number of different ways to access the data.

The simplest are webpages which give links to each of the data
products. These pages (one each for the Deep ``full version'', Deep
``best seeing version'' and Wide) also allow users to select multiple
images and catalogs for parallel download with the CADC's
DownloadManager.

There is also a graphical search tool based on Google Maps which
allows users to pan and zoom around the fields, and select images for
download. The images are previewed in this tool as RGB color images, constructed
using the \I, \R and \G-band images. This tool allows users
to select and download image cutouts.

The CFHTLS Deep fields are located at points on the sky which have
considerable coverage at other wavelengths, as discussed in Section
\ref{sec:field}.  A set of Google Sky tools, one per Deep fields are
provided which allow users to display color previews at these other
wavelengths. 

The images are large: 1.7 gigabytes or so in size. Many users require
only a fraction of the image. Therefore a cutout service is also
provided, which can be accessed either via the graphical search tool
or by a separate search page. The search page allows users to enter
coordinates and retrieve subsections of the images centered on those
coordinates.  Users can enter a single coordinate or multiple
coordinates.  The tool also allows users to search by object name. The
names are resolved to coordinates using either SIMBAD \citep{simbad}
or NED \citep{NED} and then the images are searched as before.

The merged catalogs described in section \ref{ssec:mergecat} can be
downloaded directly. They can also be queried using a system similar
to the SDSS CasJobs \citep{casjobs}. This service allows users to
construct arbitrary queries in SQL and retrieve a variety of parameters
or combination of parameters.

\acknowledgments

This research used the facilities of the Canadian Astronomy Data
Centre operated by the National Research Council of Canada with the
support of the Canadian Space Agency.

Based on observations obtained with MegaPrime/MegaCam, a joint project
of CFHT and CEA/DAPNIA, at the Canada-France-Hawaii Telescope (CFHT)
which is operated by the National Research Council (NRC) of Canada,
the Institute National des Sciences de l'Univers of the Centre
National de la Recherche Scientifique of France, and the University of
Hawaii.

This research has made use of the NASA/IPAC Extragalactic Database
(NED) which is operated by the Jet Propulsion Laboratory, California
Institute of Technology, under contract with the National Aeronautics
and Space Administration.

Funding for the SDSS and SDSS-II has been provided by the Alfred
P. Sloan Foundation, the Participating Institutions, the National
Science Foundation, the U.S. Department of Energy, the National
Aeronautics and Space Administration, the Japanese Monbukagakusho, the
Max Planck Society, and the Higher Education Funding Council for
England. The SDSS Web Site is http://www.sdss.org/.

The SDSS is managed by the Astrophysical Research Consortium for the
Participating Institutions. The Participating Institutions are the
American Museum of Natural History, Astrophysical Institute Potsdam,
University of Basel, University of Cambridge, Case Western Reserve
University, University of Chicago, Drexel University, Fermilab, the
Institute for Advanced Study, the Japan Participation Group, Johns
Hopkins University, the Joint Institute for Nuclear Astrophysics, the
Kavli Institute for Particle Astrophysics and Cosmology, the Korean
Scientist Group, the Chinese Academy of Sciences (LAMOST), Los Alamos
National Laboratory, the Max-Planck-Institute for Astronomy (MPIA),
the Max-Planck-Institute for Astrophysics (MPA), New Mexico State
University, Ohio State University, University of Pittsburgh,
University of Portsmouth, Princeton University, the United States
Naval Observatory, and the University of Washington.


{\it Facilities:} \facility{CFHT}.

\begin{deluxetable}{lcccccccccccccccc}
\label{tab:deepfull}
\tablecaption{Deep full stacks: limiting magnitudes and image qualities}
\tablehead{
\colhead{}
& \multicolumn{2}{c}{\U} 
& \multicolumn{2}{c}{\G} 
& \multicolumn{2}{c}{\R} 
& \multicolumn{2}{c}{\I} 
& \multicolumn{2}{c}{\Z}
\\
\colhead{Field} 
& \colhead{$m_{lim}$} & \colhead{IQ ($''$)}
& \colhead{$m_{lim}$} & \colhead{IQ ($''$)}
& \colhead{$m_{lim}$} & \colhead{IQ ($''$)}
& \colhead{$m_{lim}$} & \colhead{IQ ($''$)}
& \colhead{$m_{lim}$} & \colhead{IQ ($''$)}
}
\startdata
{\tt D1 }   &  27.4 &     0.94  &  27.9 &     0.89  &  27.6 &     0.83  &  27.4 &     0.78 &  26.1 &     0.81\\ 
{\tt D2 }   &  27.5 &     0.96  &  27.9 &     0.92  &  27.6 &     0.84  &  27.3 &     0.83 &  26.4 &     0.83\\ 
{\tt D3 }   &  27.4 &     0.97  &  28.0 &     0.94  &  27.8 &     0.83  &  27.6 &     0.80 &  26.4 &     0.78\\ 
{\tt D4 }   &  27.3 &     0.96  &  27.8 &     0.91  &  27.7 &     0.83  &  27.2 &     0.80 &  26.2 &     0.78\\ 
\enddata
\end{deluxetable}

\begin{deluxetable}{lcccccccccccccccc}
\label{tab:deepiq}
\tablecaption{Deep good seeing stacks: limiting magnitudes and image qualities}
\tablehead{
\colhead{}
& \multicolumn{2}{c}{\U} 
& \multicolumn{2}{c}{\G} 
& \multicolumn{2}{c}{\R} 
& \multicolumn{2}{c}{\I} 
& \multicolumn{2}{c}{\Z}
\\
\colhead{Field} 
& \colhead{$m_{lim}$} & \colhead{IQ ($''$)}
& \colhead{$m_{lim}$} & \colhead{IQ ($''$)}
& \colhead{$m_{lim}$} & \colhead{IQ ($''$)}
& \colhead{$m_{lim}$} & \colhead{IQ ($''$)}
& \colhead{$m_{lim}$} & \colhead{IQ ($''$)}
}
\startdata
{\tt D1 }   &  27.1 &     0.80  &  27.2 &     0.64  &  27.3 &     0.64  &  27.3 &     0.64  &  25.9 &     0.64\\ 
{\tt D2 }   &  27.4 &     0.80  &  27.4 &     0.65  &  27.4 &     0.64  &  27.1 &     0.64  &  26.2 &     0.64\\ 
{\tt D3 }   &  27.1 &     0.78  &  27.5 &     0.65  &  27.5 &     0.64  &  27.3 &     0.64  &  26.1 &     0.64\\ 
{\tt D4 }   &  26.8 &     0.78  &  27.0 &     0.64  &  27.0 &     0.64  &  26.8 &     0.64  &  25.9 &     0.64\\ 
\enddata
\end{deluxetable}

\begin{deluxetable}{lcccccccccccccccc}
\label{tab:wide}
\tablecaption{Wide stacks: limiting magnitudes and image qualities}
\tabletypesize{\small}
\tablehead{
\colhead{}
& \multicolumn{2}{c}{\U} 
& \multicolumn{2}{c}{\G} 
& \multicolumn{2}{c}{\R} 
& \multicolumn{2}{c}{\I} 
& \multicolumn{2}{c}{\Z}
\\
\colhead{Pointing} 
& \colhead{$m_{lim}$} & \colhead{IQ ($''$)}
& \colhead{$m_{lim}$} & \colhead{IQ ($''$)}
& \colhead{$m_{lim}$} & \colhead{IQ ($''$)}
& \colhead{$m_{lim}$} & \colhead{IQ ($''$)}
& \colhead{$m_{lim}$} & \colhead{IQ ($''$)}
}
\startdata
{\tt W1+0+0}  &  26.1 &     0.83  &  26.4 &     0.86  &  26.0 &     0.75  &  25.8 &     0.74  &  24.0 &     0.96\\ 
{\tt W1+0+1}  &  25.7 &     1.01  &  26.3 &     0.91  &  26.1 &     0.83  &  26.0 &     0.64  &  24.9 &     0.78\\ 
{\tt W1+0+2}  &  25.9 &     0.97  &  26.5 &     0.92  &  25.8 &     0.86  &  26.0 &     0.57  &  24.6 &     0.84\\ 
{\tt W1+0+3}  &  26.0 &     0.86  &  26.3 &     0.90  &  25.7 &     0.92  &  25.7 &     0.80  &  24.5 &     0.90\\ 
{\tt W1+0-1}  &  26.0 &     1.06  &  26.5 &     0.89  &  25.9 &     0.79  &  25.9 &     0.55  &  24.1 &     1.01\\ 
{\tt W1+0-2}  &  25.9 &     1.11  &  26.8 &     0.66  &  25.8 &     0.84  &  25.8 &     0.69  &  24.7 &     0.66\\ 
{\tt W1+0-3}  &  26.2 &     1.09  &  26.5 &     0.91  &  25.8 &     0.86  &  25.8 &     0.72  &  24.5 &     0.78\\ 
{\tt W1+0-4}  &  26.1 &     0.84  &  26.6 &     0.95  &  25.9 &     0.77  &  25.6 &     0.71  &  25.0 &     0.56\\ 
{\tt W1+1+0}  &  26.3 &     0.85  &  26.6 &     0.79  &  25.9 &     0.73  &  25.8 &     0.57  &  24.5 &     0.84\\ 
{\tt W1+1+1}  &  25.8 &     0.97  &  26.7 &     0.92  &  25.8 &     0.84  &  26.0 &     0.86  &  25.1 &     0.79\\ 
{\tt W1+1+2}  &  25.7 &     1.03  &  26.2 &     0.86  &  26.1 &     0.67  &  25.5 &     0.81  &  24.9 &     0.69\\ 
{\tt W1+1+3}  &  26.0 &     0.88  &  26.5 &     0.83  &  25.9 &     0.95  &  25.6 &     0.80  &  24.9 &     0.68\\ 
{\tt W1+1-1}  &  26.0 &     0.97  &  26.6 &     0.98  &  26.0 &     0.74  &  26.3 &     0.81  &  24.2 &     0.95\\ 
{\tt W1+1-2}  &  26.0 &     1.10  &  26.7 &     0.76  &  25.9 &     0.84  &  26.0 &     0.81  &  24.6 &     0.78\\ 
{\tt W1+1-3}  &  26.1 &     1.05  &  26.5 &     0.81  &  25.8 &     0.82  &  25.7 &     0.70  &  24.7 &     0.80\\ 
{\tt W1+1-4}  &  26.1 &     0.92  &  26.6 &     0.93  &  26.1 &     0.76  &  25.9 &     0.66  &  24.9 &     0.58\\ 
{\tt W1+2+0}  &  25.7 &     1.01  &  26.5 &     0.86  &  26.0 &     0.82  &  25.4 &     0.74  &  24.3 &     0.87\\ 
{\tt W1+2+1}  &  25.9 &     1.06  &  26.3 &     1.07  &  25.7 &     0.88  &  25.5 &     0.88  &  25.0 &     0.68\\ 
{\tt W1+2+2}  &  25.9 &     0.96  &  26.3 &     0.91  &  25.7 &     0.89  &  25.9 &     0.88  &  24.7 &     0.82\\ 
{\tt W1+2+3}  &  26.3 &     1.00  &  26.5 &     0.83  &  26.2 &     0.70  &  25.6 &     0.78  &  25.2 &     0.60\\ 
{\tt W1+2-1}  &  26.3 &     0.96  &  26.6 &     0.78  &  26.0 &     0.72  &  25.7 &     0.77  &  24.5 &     1.02\\ 
{\tt W1+2-2}  &  26.1 &     1.10  &  26.5 &     0.77  &  25.9 &     0.73  &  25.8 &     0.69  &  24.2 &     1.00\\ 
{\tt W1+2-3}  &  26.3 &     0.81  &  26.6 &     0.73  &  25.8 &     0.81  &  25.5 &     0.73  &  24.4 &     0.92\\ 
{\tt W1+2-4}  &  26.1 &     0.89  &  26.7 &     0.85  &  25.8 &     0.77  &  25.7 &     0.75  &  24.9 &     0.61\\ 
{\tt W1+3+0}  &  26.3 &     0.64  &  26.3 &     0.92  &  25.9 &     0.83  &  25.7 &     0.73  &  24.6 &     0.86\\ 
{\tt W1+3+1}  &  26.1 &     0.90  &  26.4 &     0.99  &  25.8 &     0.82  &  25.8 &     0.87  &  24.5 &     0.82\\ 
{\tt W1+3+2}  &  26.2 &     1.10  &  26.3 &     0.87  &  25.9 &     0.83  &  25.6 &     0.70  &  24.6 &     0.64\\ 
{\tt W1+3+3}  &  25.7 &     1.25  &  26.3 &     0.90  &  25.7 &     0.82  &  25.3 &     0.87  &  24.7 &     0.62\\ 
{\tt W1+3-1}  &  25.9 &     1.05  &  26.5 &     0.80  &  26.1 &     0.74  &  25.8 &     0.77  &  24.8 &     0.74\\ 
{\tt W1+3-2}  &  26.4 &     0.85  &  26.6 &     0.74  &  25.8 &     0.74  &  25.8 &     0.72  &  24.6 &     0.82\\ 
{\tt W1+3-3}  &  26.2 &     0.80  &  26.7 &     0.67  &  25.8 &     0.87  &  25.7 &     0.64  &  24.5 &     0.90\\ 
{\tt W1+3-4}  &  26.0 &     1.04  &  26.3 &     0.90  &  25.8 &     0.78  &  26.1 &     0.66  &  24.8 &     0.63\\ 
{\tt W1+4+0}  &  26.4 &     0.75  &  26.5 &     0.91  &  25.9 &     0.74  &  25.3 &     0.70  &  24.7 &     0.75\\ 
{\tt W1+4+1}  &  26.3 &     0.84  &  26.4 &     0.87  &  25.7 &     0.76  &  25.3 &     0.84  &  25.0 &     0.49\\ 
{\tt W1+4+2}  &  26.3 &     0.79  &  26.7 &     0.90  &  25.9 &     0.67  &  25.6 &     0.94  &  24.8 &     0.61\\ 
{\tt W1+4+3}  &  26.0 &     1.00  &  26.4 &     0.89  &  25.8 &     0.78  &  25.2 &     0.88  &  24.2 &     0.86\\ 
{\tt W1+4-1}  &  25.9 &     1.17  &  26.6 &     0.89  &  25.9 &     0.77  &  26.0 &     0.61  &  24.5 &     0.85\\ 
{\tt W1+4-2}  &  26.4 &     0.74  &  26.6 &     0.81  &  25.9 &     0.73  &  25.9 &     0.73  &  24.4 &     0.75\\ 
{\tt W1+4-3}  &  26.0 &     0.94  &  26.5 &     0.85  &  25.7 &     0.92  &  25.8 &     0.73  &  24.5 &     0.79\\ 
{\tt W1+4-4}  &  26.2 &     0.91  &  26.1 &     1.00  &  26.1 &     0.92  &  26.0 &     0.83  &  24.8 &     0.62\\ 
{\tt W1-1+0}  &  26.0 &     0.97  &  26.5 &     0.73  &  26.0 &     0.74  &  25.7 &     0.64  &  24.3 &     0.87\\ 
{\tt W1-1+1}  &  26.1 &     1.06  &  26.7 &     1.02  &  25.8 &     0.83  &  25.9 &     0.62  &  24.5 &     0.79\\ 
{\tt W1-1+2}  &  25.9 &     0.96  &  26.5 &     0.92  &  25.9 &     0.88  &  25.7 &     0.64  &  25.0 &     0.72\\ 
{\tt W1-1+3}  &  26.1 &     0.80  &  26.4 &     1.00  &  25.8 &     0.89  &  25.6 &     0.81  &  24.9 &     0.66\\ 
{\tt W1-1-1}  &  26.7 &     0.70  &  26.9 &     0.59  &  25.7 &     0.90  &  25.9 &     0.58  &  24.1 &     0.91\\ 
{\tt W1-1-2}  &  26.3 &     0.83  &  26.9 &     0.62  &  26.0 &     0.75  &  25.6 &     0.86  &  24.1 &     1.03\\ 
{\tt W1-1-3}  &  26.3 &     0.76  &  26.2 &     1.01  &  25.8 &     0.77  &  25.7 &     0.71  &  24.5 &     1.01\\ 
{\tt W1-1-4}  &  26.4 &     0.70  &  26.7 &     1.05  &  25.9 &     0.76  &  25.5 &     0.88  &  24.8 &     0.68\\ 
{\tt W1-2+0}  &  26.5 &     0.82  &  26.8 &     0.64  &  25.8 &     0.84  &  26.1 &     0.52  &  24.4 &     0.96\\ 
{\tt W1-2+1}  &  26.4 &     0.81  &  26.8 &     0.72  &  25.9 &     0.75  &  25.4 &     0.89  &  24.9 &     0.76\\ 
{\tt W1-2+2}  &  26.4 &     0.79  &  26.2 &     0.92  &  25.9 &     0.73  &  25.2 &     0.82  &  24.4 &     0.88\\ 
{\tt W1-2+3}  &  26.2 &     0.83  &  26.1 &     0.90  &  26.0 &     0.74  &  25.3 &     0.76  &  24.4 &     0.67\\ 
{\tt W1-2-1}  &  26.3 &     0.79  &  26.9 &     0.66  &  25.9 &     0.86  &  26.1 &     0.50  &  24.5 &     1.07\\ 
{\tt W1-2-2}  &  26.3 &     0.79  &  26.6 &     0.70  &  26.1 &     0.76  &  25.9 &     0.53  &  24.8 &     0.62\\ 
{\tt W1-2-3}  &  26.4 &     0.72  &  26.3 &     0.81  &  25.8 &     0.83  &  25.8 &     0.65  &  24.2 &     1.05\\ 
{\tt W1-2-4}  &  26.3 &     0.92  &  26.5 &     0.96  &  26.0 &     0.75  &  26.0 &     0.58  &  25.2 &     0.55\\ 
{\tt W1-3+0}  &  26.0 &     0.88  &  26.3 &     0.98  &  25.9 &     0.85  &  25.6 &     0.78  &  24.7 &     0.65\\ 
{\tt W1-3+1}  &  26.3 &     0.90  &  26.3 &     0.86  &  25.9 &     0.85  &  25.6 &     0.79  &  24.7 &     0.59\\ 
{\tt W1-3+2}  &  26.3 &     0.86  &  26.1 &     0.98  &  25.7 &     0.89  &  25.7 &     0.77  &  24.8 &     0.66\\ 
{\tt W1-3+3}  &  26.1 &     0.85  &  26.7 &     0.75  &  25.8 &     0.68  &  25.7 &     0.74  &  24.6 &     0.67\\ 
{\tt W1-3-1}  &  25.9 &     1.01  &  26.5 &     0.80  &  25.9 &     0.97  &  25.6 &     0.82  &  24.9 &     0.52\\ 
{\tt W1-3-2}  &  25.8 &     1.10  &  26.6 &     0.69  &  25.8 &     0.83  &  25.5 &     0.75  &  24.7 &     0.74\\ 
{\tt W1-3-3}  &  26.0 &     0.74  &  26.6 &     0.70  &  25.9 &     0.75  &  25.8 &     0.75  &  24.9 &     0.73\\ 
{\tt W1-3-4}  &  26.3 &     0.97  &  26.4 &     0.89  &  26.1 &     0.78  &  25.8 &     0.65  &  25.0 &     0.55\\ 
{\tt W1-4+0}  &  25.9 &     1.05  &  26.2 &     0.90  &  25.8 &     0.90  &  26.0 &     0.61  &  24.5 &     0.74\\ 
{\tt W1-4+1}  &  25.7 &     1.09  &  26.2 &     0.98  &  25.7 &     0.88  &  25.7 &     0.71  &  24.7 &     0.74\\ 
{\tt W1-4+2}  &  25.8 &     1.10  &  26.6 &     1.05  &  25.8 &     0.77  &  25.6 &     0.68  &  24.5 &     0.88\\ 
{\tt W1-4+3}  &  26.2 &     0.85  &  26.3 &     0.85  &  26.0 &     0.72  &  25.5 &     0.74  &  24.3 &     0.94\\ 
{\tt W1-4-1}  &  26.4 &     0.87  &  26.4 &     0.77  &  25.7 &     0.94  &  25.6 &     0.73  &  24.0 &     1.03\\ 
{\tt W1-4-2}  &  26.5 &     0.84  &  26.2 &     0.93  &  25.6 &     1.02  &  25.7 &     0.70  &  24.6 &     1.06\\ 
{\tt W1-4-3}  &  26.0 &     0.88  &  26.5 &     0.92  &  25.5 &     1.01  &  25.5 &     0.69  &  24.4 &     0.89\\ 
{\tt W1-4-4}  &  26.1 &     0.81  &  26.5 &     0.84  &  25.5 &     0.90  &  25.8 &     0.90  &  24.6 &     0.64\\ 
{\tt W2+0+0}  &  25.9 &     0.96  &  26.6 &     0.90  &  25.9 &     0.72  &  25.7 &     0.77  &  24.3 &     0.91\\ 
{\tt W2+0+1}  &  26.0 &     0.94  &  26.2 &     0.88  &  25.7 &     0.88  &  25.7 &     0.61  &  24.6 &     0.55\\ 
{\tt W2+0+2}  &  25.7 &     1.08  &  26.4 &     0.91  &  26.0 &     0.63  &  25.9 &     0.73  &  24.4 &     0.93\\ 
{\tt W2+0+3}  &  26.1 &     0.81  &  26.2 &     0.74  &  25.9 &     0.83  &  25.6 &     0.56  &  24.2 &     0.84\\ 
{\tt W2+0-1}  &  25.9 &     1.20  &  26.1 &     0.98  &  25.7 &     0.93  &  25.7 &     0.66  &  24.5 &     1.09\\ 
{\tt W2+1+0}  &  26.0 &     0.79  &  26.6 &     0.85  &  25.6 &     0.92  &  25.6 &     0.77  &  24.6 &     0.89\\ 
{\tt W2+1+1}  &  25.9 &     0.99  &  26.4 &     0.81  &  25.9 &     0.91  &  25.7 &     0.86  &  24.7 &     0.85\\ 
{\tt W2+1+2}  &  25.9 &     1.25  &  26.2 &     0.92  &  26.0 &     0.87  &  25.5 &     0.83  &  24.5 &     0.70\\ 
{\tt W2+1+3}  &  25.9 &     1.08  &  26.2 &     0.91  &  26.0 &     0.81  &  25.4 &     0.88  &  24.5 &     0.77\\ 
{\tt W2+1-1}  &  26.3 &     0.83  &  26.0 &     0.92  &  25.5 &     1.05  &  25.5 &     0.80  &  24.6 &     0.74\\ 
{\tt W2+2+0}  &  25.9 &     0.97  &  26.2 &     0.83  &  25.7 &     0.81  &  25.7 &     0.78  &  24.4 &     1.01\\ 
{\tt W2+2+1}  &  26.0 &     1.01  &  26.3 &     0.78  &  25.9 &     0.79  &  25.8 &     0.58  &  24.7 &     0.81\\ 
{\tt W2+2+2}  &  26.1 &     1.19  &  26.3 &     0.80  &  26.5 &     0.63  &  26.0 &     0.84  &  24.6 &     0.79\\ 
{\tt W2+2+3}  &  26.0 &     0.88  &  26.9 &     0.66  &  26.0 &     0.81  &  25.4 &     0.86  &  24.5 &     0.93\\ 
{\tt W2+2-1}  &  26.4 &     0.75  &  26.3 &     0.94  &  25.7 &     0.78  &  25.6 &     0.78  &  24.6 &     0.81\\ 
{\tt W2+3+0}  &  25.9 &     0.92  &  26.3 &     0.85  &  25.6 &     0.99  &  25.9 &     0.60  &  24.9 &     0.82\\ 
{\tt W2+3+1}  &  26.3 &     0.88  &  26.3 &     0.93  &  26.0 &     0.75  &  25.2 &     0.89  &  24.5 &     0.79\\ 
{\tt W2+3+2}  &  26.1 &     1.02  &  26.3 &     0.97  &  25.9 &     0.80  &  25.5 &     0.80  &  25.0 &     0.76\\ 
{\tt W2+3+3}  &  26.4 &     0.84  &  26.4 &     0.99  &  26.1 &     0.85  &  25.5 &     0.65  &  24.5 &     0.90\\ 
{\tt W2+3-1}  &  26.3 &     1.01  &  26.2 &     0.86  &  25.7 &     0.82  &  25.4 &     0.83  &  24.7 &     0.64\\ 
{\tt W2-1+0}  &  25.8 &     1.01  &  27.0 &     0.58  &  25.9 &     0.73  &  26.0 &     0.57  &  24.6 &     0.80\\ 
{\tt W2-1+1}  &  26.0 &     0.89  &  26.5 &     1.02  &  25.9 &     0.94  &  25.9 &     0.68  &  24.3 &     0.89\\ 
{\tt W2-1+2}  &  25.9 &     1.16  &  26.6 &     0.80  &  25.7 &     0.84  &  25.8 &     0.55  &  24.6 &     0.80\\ 
{\tt W2-1+3}  &  25.7 &     1.10  &  26.7 &     0.76  &  26.3 &     0.67  &  25.7 &     0.73  &  24.5 &     0.77\\ 
{\tt W2-1-1}  &  25.6 &     0.96  &  26.5 &     0.98  &  25.8 &     0.84  &  25.8 &     0.69  &  24.6 &     0.78\\ 
{\tt W3+0+0}  &  25.6 &     1.03  &  26.3 &     0.96  &  25.5 &     0.95  &  25.3 &     0.83  &  24.2 &     0.81\\ 
{\tt W3+0+1}  &  25.9 &     0.99  &  26.4 &     0.90  &  25.6 &     0.91  &  25.7 &     0.72  &  24.7 &     0.77\\ 
{\tt W3+0+2}  &  25.6 &     1.06  &  26.4 &     0.85  &  25.9 &     0.79  &  25.6 &     0.71  &  24.4 &     0.85\\ 
{\tt W3+0+3}  &  26.0 &     0.96  &  26.4 &     0.84  &  26.1 &     0.60  &  25.5 &     0.64  &  24.8 &     0.92\\ 
{\tt W3+0-1}  &  25.6 &     0.99  &  26.0 &     1.08  &  25.8 &     1.13  &  25.9 &     0.72  &  25.2 &     0.80\\ 
{\tt W3+0-2}  &  26.3 &     0.82  &  26.4 &     0.84  &  25.6 &     1.00  &  25.5 &     0.69  &  24.6 &     0.75\\ 
{\tt W3+0-3}  &  26.1 &     0.75  &  26.2 &     0.91  &  25.9 &     0.92  &  25.1 &     0.80  &  24.8 &     0.86\\ 
{\tt W3+1+0}  &  26.3 &     0.82  &  26.4 &     0.87  &  25.8 &     0.90  &  25.4 &     0.75  &  24.6 &     0.99\\ 
{\tt W3+1+1}  &  26.0 &     0.95  &  26.5 &     0.83  &  25.7 &     0.89  &  25.7 &     0.77  &  24.7 &     0.77\\ 
{\tt W3+1+2}  &  26.5 &     0.88  &  26.6 &     0.72  &  26.0 &     0.73  &  25.9 &     0.69  &  24.5 &     0.78\\ 
{\tt W3+1+3}  &  26.4 &     0.84  &  26.3 &     0.87  &  26.1 &     0.77  &  25.7 &     0.70  &  24.7 &     0.82\\ 
{\tt W3+1-1}  &  25.6 &     1.16  &  26.4 &     0.90  &  25.7 &     0.92  &  25.6 &     0.72  &  24.6 &     1.00\\ 
{\tt W3+1-2}  &  25.8 &     1.17  &  26.5 &     0.83  &  25.5 &     0.93  &  25.4 &     0.79  &  24.5 &     0.84\\ 
{\tt W3+1-3}  &  25.7 &     1.16  &  26.7 &     1.02  &  26.0 &     0.87  &  25.7 &     0.72  &  24.6 &     0.83\\ 
{\tt W3+2+0}  &  26.0 &     0.91  &  26.7 &     0.89  &  25.8 &     0.85  &  25.6 &     0.76  &  24.5 &     0.64\\ 
{\tt W3+2+1}  &  26.3 &     0.76  &  26.6 &     0.97  &  26.0 &     0.72  &  25.7 &     0.85  &  24.4 &     0.83\\ 
{\tt W3+2+2}  &  26.6 &     0.65  &  26.4 &     0.83  &  25.8 &     0.81  &  25.7 &     0.60  &  24.5 &     0.65\\ 
{\tt W3+2+3}  &  26.0 &     1.00  &  26.5 &     0.74  &  26.0 &     0.66  &  25.6 &     0.73  &  25.0 &     0.72\\ 
{\tt W3+2-1}  &  26.2 &     0.90  &  26.4 &     0.94  &  25.9 &     1.03  &  25.7 &     0.74  &  24.3 &     0.74\\ 
{\tt W3+2-2}  &  26.3 &     0.84  &  26.6 &     0.89  &  25.5 &     1.02  &  25.9 &     0.64  &  24.7 &     0.61\\ 
{\tt W3+2-3}  &  26.3 &     0.87  &  26.6 &     0.97  &  25.6 &     0.99  &  25.5 &     0.67  &  24.7 &     0.61\\ 
{\tt W3+3+0}  &  26.0 &     1.02  &  26.3 &     1.02  &  26.0 &     0.79  &  25.8 &     0.69  &  24.6 &     0.82\\ 
{\tt W3+3+1}  &  26.1 &     0.91  &  26.4 &     1.01  &  25.9 &     0.80  &  25.5 &     0.90  &  24.3 &     0.73\\ 
{\tt W3+3+2}  &  25.9 &     0.96  &  26.3 &     0.94  &  26.1 &     0.71  &  25.6 &     0.81  &  24.2 &     0.74\\ 
{\tt W3+3+3}  &  25.9 &     0.95  &  26.3 &     0.72  &  26.1 &     0.71  &  25.5 &     0.81  &  25.3 &     0.68\\ 
{\tt W3+3-1}  &  26.1 &     0.90  &  26.7 &     0.91  &  25.9 &     0.83  &  25.7 &     0.94  &  24.7 &     0.71\\ 
{\tt W3+3-2}  &  26.3 &     1.01  &  26.6 &     1.02  &  25.9 &     0.85  &  25.6 &     0.67  &  24.8 &     0.66\\ 
{\tt W3+3-3}  &  26.3 &     0.89  &  26.6 &     0.95  &  25.9 &     0.85  &  25.6 &     0.82  &  24.7 &     0.72\\ 
{\tt W3-1+0}  &  26.1 &     0.71  &  26.3 &     1.07  &  26.0 &     0.75  &  25.4 &     0.60  &  24.1 &     0.89\\ 
{\tt W3-1+1}  &  26.1 &     0.89  &  26.5 &     0.78  &  25.7 &     0.89  &  25.5 &     0.85  &  24.4 &     0.88\\ 
{\tt W3-1+2}  &  25.7 &     1.03  &  26.7 &     0.77  &  25.8 &     0.87  &  25.6 &     0.72  &  24.5 &     0.78\\ 
{\tt W3-1+3}  &  26.1 &     0.98  &  26.3 &     0.89  &  25.7 &     0.82  &  26.1 &     0.89  &  24.6 &     0.68\\ 
{\tt W3-1-1}  &  25.9 &     1.03  &  26.7 &     0.80  &  25.6 &     0.83  &  25.6 &     0.60  &  24.3 &     0.88\\ 
{\tt W3-1-2}  &  25.3 &     0.90  &  26.3 &     0.91  &  26.0 &     0.75  &  25.4 &     0.72  &  24.6 &     0.73\\ 
{\tt W3-1-3}  &  25.7 &     0.77  &  26.5 &     0.88  &  25.5 &     0.90  &  25.7 &     0.72  &  24.7 &     0.66\\ 
{\tt W3-2+0}  &  26.2 &     0.72  &  26.5 &     0.75  &  26.2 &     0.73  &  25.1 &     0.82  &  24.8 &     0.62\\ 
{\tt W3-2+1}  &  26.1 &     0.88  &  26.6 &     0.89  &  25.8 &     0.91  &  25.7 &     0.87  &  24.7 &     0.74\\ 
{\tt W3-2+2}  &  25.9 &     0.94  &  26.8 &     0.68  &  25.8 &     0.86  &  25.6 &     0.63  &  24.5 &     0.88\\ 
{\tt W3-2+3}  &  26.2 &     0.89  &  26.5 &     0.86  &  26.1 &     0.78  &  26.0 &     0.90  &  25.1 &     0.77\\ 
{\tt W3-2-1}  &  25.8 &     0.95  &  26.1 &     1.02  &  25.9 &     0.70  &  26.0 &     0.72  &  24.8 &     0.60\\ 
{\tt W3-2-2}  &  26.1 &     0.79  &  26.3 &     0.94  &  25.9 &     0.80  &  25.6 &     0.71  &  24.6 &     0.71\\ 
{\tt W3-2-3}  &  25.9 &     1.22  &  26.4 &     0.99  &  25.8 &     0.84  &  25.5 &     0.76  &  24.2 &     0.90\\ 
{\tt W3-3+0}  &  26.1 &     0.75  &  26.5 &     0.77  &  26.1 &     0.67  &  25.8 &     0.58  &  24.7 &     0.68\\ 
{\tt W3-3+1}  &  26.3 &     0.70  &  26.8 &     0.78  &  25.7 &     0.92  &  26.1 &     0.92  &  24.7 &     0.69\\ 
{\tt W3-3+2}  &  25.8 &     1.08  &  26.7 &     0.70  &  26.1 &     0.88  &  25.4 &     0.72  &  24.9 &     0.64\\ 
{\tt W3-3+3}  &  26.2 &     1.05  &  26.5 &     0.81  &  26.0 &     0.75  &  25.6 &     0.71  &  24.9 &     0.75\\ 
{\tt W3-3-1}  &  25.6 &     1.00  &  26.0 &     0.88  &  25.8 &     0.74  &  25.6 &     0.79  &  24.8 &     0.63\\ 
{\tt W3-3-2}  &  26.1 &     0.67  &  26.5 &     0.69  &  25.6 &     0.78  &  25.7 &     0.58  &  24.8 &     0.59\\ 
{\tt W3-3-3}  &  26.1 &     0.74  &  26.3 &     0.84  &  25.5 &     0.96  &  25.8 &     0.54  &  24.8 &     0.64\\ 
{\tt W4+0+0}  &  25.7 &     1.14  &  26.3 &     0.85  &  26.1 &     0.63  &  25.6 &     0.74  &  25.0 &     0.74\\ 
{\tt W4+0+1}  &  25.9 &     0.99  &  26.4 &     0.97  &  26.1 &     0.62  &  25.5 &     0.70  &  24.6 &     0.60\\ 
{\tt W4+0-1}  &  26.2 &     0.80  &  26.7 &     0.86  &  26.0 &     0.75  &  25.9 &     0.59  &  24.7 &     0.56\\ 
{\tt W4+0-2}  &  26.3 &     0.75  &  26.8 &     0.79  &  25.9 &     0.75  &  25.9 &     0.66  &  24.8 &     0.70\\ 
{\tt W4+1+0}  &  25.9 &     0.91  &  26.4 &     0.70  &  25.8 &     0.75  &  25.7 &     0.56  &  24.7 &     0.62\\ 
{\tt W4+1+1}  &  26.2 &     0.80  &  26.8 &     0.60  &  25.8 &     0.77  &  25.9 &     0.59  &  24.7 &     0.60\\ 
{\tt W4+1-1}  &  25.9 &     0.92  &  26.1 &     0.90  &  25.9 &     0.71  &  26.0 &     0.67  &  24.5 &     0.71\\ 
{\tt W4+1-2}  &  25.9 &     0.91  &  26.2 &     0.89  &  26.0 &     0.67  &  25.5 &     0.80  &  24.8 &     0.56\\ 
{\tt W4+2+0}  &  25.9 &     0.85  &  26.3 &     0.73  &  26.0 &     0.69  &  25.9 &     0.63  &  24.1 &     0.86\\ 
{\tt W4+2-1}  &  25.5 &     1.07  &  26.3 &     0.83  &  25.8 &     0.72  &  25.6 &     0.73  &  24.9 &     0.79\\ 
{\tt W4+2-2}  &  25.7 &     1.15  &  26.3 &     0.79  &  25.8 &     0.68  &  26.0 &     0.76  &  24.8 &     0.82\\ 
{\tt W4-1+0}  &  25.7 &     0.72  &  26.5 &     0.90  &  26.0 &     0.70  &  25.7 &     0.71  &  24.4 &     0.81\\ 
{\tt W4-1+1}  &  26.1 &     0.80  &  26.8 &     0.86  &  25.9 &     0.69  &  25.9 &     0.87  &  24.6 &     0.74\\ 
{\tt W4-1+2}  &  26.2 &     0.83  &  26.7 &     0.71  &  25.8 &     0.66  &  25.9 &     0.58  &  24.3 &     1.14\\ 
{\tt W4-1+3}  &  26.0 &     0.97  &  26.5 &     0.85  &  25.7 &     0.83  &  26.0 &     0.55  &  24.1 &     0.94\\ 
{\tt W4-1-1}  &  26.1 &     0.75  &  26.9 &     0.90  &  25.8 &     0.78  &  26.1 &     0.77  &  24.8 &     0.55\\ 
{\tt W4-1-2}  &  26.5 &     0.64  &  26.6 &     0.87  &  25.9 &     0.84  &  25.6 &     0.80  &  24.9 &     0.55\\ 
{\tt W4-2+0}  &  25.9 &     1.01  &  26.5 &     0.78  &  25.8 &     0.82  &  25.9 &     0.80  &  24.6 &     0.68\\ 
{\tt W4-2+1}  &  25.4 &     1.09  &  26.0 &     0.91  &  26.1 &     0.66  &  25.2 &     0.62  &  24.4 &     0.87\\ 
{\tt W4-2+2}  &  25.9 &     1.01  &  26.5 &     0.80  &  25.7 &     0.73  &  25.6 &     0.74  &  24.2 &     0.83\\ 
{\tt W4-2+3}  &  26.2 &     0.85  &  26.5 &     0.88  &  25.6 &     0.91  &  25.7 &     0.78  &  24.3 &     0.69\\ 
{\tt W4-3+0}  &  26.0 &     0.99  &  26.7 &     0.74  &  25.8 &     0.76  &  25.4 &     0.56  &  24.5 &     0.87\\ 
{\tt W4-3+1}  &  26.1 &     1.06  &  26.6 &     0.78  &  25.8 &     0.73  &  25.6 &     0.48  &  24.3 &     0.87\\ 
{\tt W4-3+2}  &  26.0 &     0.89  &  26.4 &     0.82  &  25.7 &     0.66  &  25.1 &     0.74  &  24.5 &     0.77\\ 
{\tt W4-3+3}  &  26.0 &     0.93  &  26.6 &     1.06  &  25.6 &     0.83  &  25.9 &     0.63  &  24.5 &     0.71\\ 

\enddata
\end{deluxetable}

\begin{table}
\caption{Sextractor detection parameters}
\label{tab:sex}
\begin{tabular}{lr}
\tableline\tableline
Parameter & Value\\
\tableline
{\tt DETECT\underline{~}TYPE     }& {\tt CCD             }\\
{\tt DETECT\underline{~}MINAREA  }& {\tt 3               }\\
{\tt DETECT\underline{~}THRESH   }& {\tt 1               }\\
{\tt ANALYSIS\underline{~}THRESH }& {\tt 1               }\\
{\tt THRESH\underline{~}TYPE     }& {\tt RELATIVE        }\\
{\tt FILTER                      }& {\tt Y               }\\
{\tt FILTER\underline{~}NAME     }& {\tt gauss\underline{~}3.0\underline{~}7x7.conv }\\
{\tt DEBLEND\underline{~}NTHRESH }& {\tt 32              }\\
{\tt DEBLEND\underline{~}MINCONT }& {\tt 0.002           }\\
{\tt CLEAN                       }& {\tt Y               }\\
{\tt CLEAN\underline{~}PARAM     }& {\tt 1.0             }\\

\tableline
\end{tabular}
\end{table}

\clearpage

\clearpage

\clearpage

\clearpage

\clearpage

\clearpage

\clearpage

\clearpage

\clearpage

\clearpage

\clearpage

\clearpage

\clearpage

\clearpage

\clearpage

\clearpage

\clearpage

\clearpage

\clearpage

\end{document}